\documentclass[prd,twocolumn,nofootinbib,superscriptaddress]{revtex4-2}

\usepackage{bm}
\usepackage{amsmath}
\usepackage{amssymb}
\usepackage{graphicx}
\usepackage{subfigure}
\usepackage{hyperref}
\usepackage{color}
\usepackage{comment}
\usepackage{ulem}
\usepackage{CJK}
\hypersetup{
	colorlinks=true,
	linkcolor=red,
	citecolor=blue,
}

\allowdisplaybreaks[2]

\usepackage{graphicx}
\usepackage{dcolumn}
\usepackage{bm}

\begin{document}

\preprint{APS/123-QED}

\title{Phenomenological relationship between eccentric and quasi-circular orbital binary black hole waveform}

\author{Hao Wang}
\email{husthaowang@hust.edu.cn}
\affiliation{Department of Astronomy, School of Physics, Huazhong University of Science and Technology, Wuhan 430074, China}

\author{Yuan-Chuan Zou}
\email{zouyc@hust.edu.cn}
\affiliation{Department of Astronomy, School of Physics, Huazhong University of Science and Technology, Wuhan 430074, China}

\author{Yu Liu}%
\email{yul@hust.edu.cn}
\affiliation{Department of Astronomy, School of Physics, Huazhong University of Science and Technology, Wuhan 430074, China}

\date{\today}

\begin{abstract}
Eccentricity has become an increasingly important parameter in gravitational wave studies as it can clearly reflect the dynamics of compact object mergers. Obtaining an accurate and fast gravitational waveform template is of paramount importance for accurately estimating gravitational wave parameters. This paper aims to conduct an extensive study of the phenomenological fitting model proposed by Setyawati and Ohme  \cite{Setyawati:2021gom} for adding eccentricity into quasi-circular orbital waveforms. We expand the scope of this research by studying the waveform for a mass ratio range of $[1, 7]$, an initial eccentricity range of $[0, 0.4]$, and a continuous time period beyond the fixed time period of $[-1500M, -29M]$. We also investigate the model in higher-order harmonic modes, as well as spin-aligned and spin-precessing waveforms. After expanding some fitting parameters, we have discovered that the model can be applied to mass ratios $q\in[1, 7]$. Additionally, it can be applied to almost the entire time period of numerical relativity, including up to $12000M$ prior to merger. It can accommodate higher eccentricities up to $e_0 = 0.4$, but its accuracy decreases with increasing initial eccentricity. For a specific initial eccentricity $e_0$ and time period such as $[-2000M, -300M]$, mismatchs obtained are approximately less than $10^{-4}$ for $e_0 \in [0, 0.1]$, less than $10^{-3}$ for $e_0 \in [0.1, 0.2]$, less than $10^{-2}$ for $e_0 \in [0.2, 0.3]$, and less than $10^{-1}$ for $e_0 \in [0.3, 0.4]$ for mass ratio 1-3, and an order of magnitude worse for mass ratio 4-7. The dependence of the mismatch on eccentricity is due to the fact that as the initial eccentricity increases, the eccentricity estimator $e_X$ deviates further from the expected \textit{cosine} function, leading to a larger deviation in the morphology of the eccentric waveform and a reduced accuracy in the model's fitting. It can be applied to higher-order modes and yields similar overlap results. Furthermore, by introducing a shift parameter $g$, it can be approximately applied to spin-aligned waveforms. After obtaining spin-precession effects for the special case of strong precession, our model can also be applied to the general spin-precessing case. In summary, this phenomenological model allows for the construction of eccentric gravitational wave templates for non-spinning, spin-aligned or spin-precessing binary systems. It provides an efficient method for generating templates and sheds light on the \textit{phenomenological and universal} relationship between eccentric and quasi-circular waveforms.
\end{abstract}
\maketitle

\section{Introduction}

Since the first detection of the binary black hole merger event GW150914 in 2015, the field of gravitational wave astronomy has entered a new era \cite{LIGOScientific:2016aoc}. To date, ground-based gravitational wave detectors LIGO \cite{LIGOScientific:2014qfs}, Virgo \cite{VIRGO:2014yos}, and KAGRA \cite{KAGRA:2018plz} have detected 93 gravitational wave events, including binary black holes (BHHs), two black hole-neutron star (BHNS) mergers, and two binary neutron star (NSNS) mergers \cite{LIGOcollaboration}.

Currently, gravitational wave signals are extracted using circular orbital waveform templates, as it is generally assumed that the evolution of isolated binary stars circularizes due to gravitational wave radiation. Thus, the eccentricity of the binary system is expected to be negligible when it enters the gravitational wave detection frequency band, at around 10 Hz \cite{Peters:1964zz,Peters:1963ux,Hinder:2007qu,Lidov:1976qhg}. 
However, there are several ways in which BBHs can gain eccentricity before merging. In dense regions of stars such as globular clusters \cite{Miller:2002pg,Gultekin:2005fd,OLeary:2005vqo,Rodriguez:2015oxa,Samsing:2017xmd,Rodriguez:2017pec,Rodriguez:2018pss} and galactic nuclei \cite{Gondan:2020svr,Antonini:2012ad}, BBHs can acquire eccentricity due to double-single \cite{Samsing:2013kua}, double-double interactions \cite{Zevin:2018kzq}, and gravitational capture \cite{Gondan:2020svr}. In addition, in a three-body system, such as binary objects in the vicinity of a supermassive black hole, the eccentricity of the inner binary objects will oscillate due to the Kozai-Lidov mechanism \cite{VanLandingham:2016ccd,Silsbee:2016djf,Blaes:2002cs}, which will be detectable once they enter the detection frequency band.

Gravitational waves of BBHs mergers in globular clusters entering the LIGO sensitive band, 10\% of them still maintain an eccentricity more than 0.1 according to Refs. \cite{Samsing:2017xmd,Rodriguez:2017pec}. GW190521 \cite{LIGOScientific:2020iuh} is considered to be possible a BBH merger with high mass and high eccentricity $e=0.69_{-0.22}^{+0.17}$ through 611 numerical relativity simulations \cite{Gayathri:2020coq}. With the improvement of detector sensitivity, more and more eccentric BBHs mergers will be detected by next-generation ground-based gravitational-wave detectors Einstein Telescope (ET) \cite{Punturo:2010zz} or Cosmic Explorer (CE) \cite{Reitze:2019iox}.

Errors may exist or signal-to-noise ratio will be reduced when using circular orbital waveforms for parameter estimation \cite{Chattaraj:2022tay,Brown:2009ng}. 
At present, there are some numerical relativistic simulations of eccentric BBHs mergers \cite{Habib:2019cui,Huerta:2019oxn,Boyle:2019kee,Healy:2022wdn}. 
Parameter estimation of gravitational waves requires millions of waveform templates. In general, full nonlinear numerical relativity (NR) simulation yields the most precise gravitational waveform,  but each NR simulation takes several weeks and months and is computationally expensive. 

Numerous analytical gravitational waveforms based on the post-Newtonian (PN) approximation have been developed in various studies, such as those referenced in \cite{Memmesheimer:2004cv,Konigsdorffer:2006zt,Boetzel:2017zza,Loutrel:2017fgu,Tanay:2016zog,Yunes:2009yz,Huerta:2014eca,Tiwari:2019jtz,Klein:2018ybm,Konigsdorffer:2006zt,Moore:2016qxz,Moore:2019xkm,Cornish:2010cd,Gopakumar:2011zz,Barreto:2014xpa,Tiwari:2020hsu}, or based on the effective-one-body (EOB) method in studies such as those cited in \cite{Ramos-Buades:2021adz,Hinderer:2017jcs,Chiaramello:2020ehz,Nagar:2021gss,Khalil:2021txt}. \citet{Islam:2021mha} developed a waveform based on the calibration of full numerical relativity.
State-of-the-art surrogate models of full inspiral-merger-ringdown (IMR) eccentric gravitational waveforms have also been developed using a hybrid of PN and numerical relativity (NR), as described in studies such as \cite{Hinder:2017sxy,Huerta:2016rwp,Ramos-Buades:2019uvh,Huerta:2017kez}, or based on the EOBNR method, as shown in studies such as \cite{Nagar:2021gss,Nagar:2020xsk,Khalil:2021txt,Cao:2017ndf,Liu:2019jpg,Yun:2021jnh,Chiaramello:2020ehz}. Special cases, such as eccentric extreme-mass-ratio inspirals (EMRIs) \cite{Munna:2019fjz,Zhang:2020rxy,McCart:2021upc,Forseth:2015oua,Munna:2020som,Tsukamoto:2021caq}, and gravitational wave bursts with high eccentricity \cite{Loutrel:2019kky}, have also been studied.

However, full numerical relativity simulations of binary black hole mergers with eccentricity are rare and not publicly available, which makes it both convenient and necessary to develop phenomenological models that can generate fast and accurate numerically relativistic eccentric gravitational waveforms.

The main objective of this paper is to perform an extensive investigation of the phenomenological model proposed by \citet{Setyawati:2021gom} that converts a circular orbital waveform into an eccentric orbital waveform. This model is capable of rapidly and easily producing a full numerical relativistic eccentric orbital waveform based on the corresponding circular orbital waveform. In addition to the eccentric waveform from the Simulating eXtreme Spacetime (SXS) catalog, we include some eccentric waveforms from the Rochester Institute of Technology (RIT) catalog, which extends the range of mass ratio to $q\in[1,7]$ and the initial eccentricity range to $e_0\in[0,0.4]$. Furthermore, we extend the time range of the waveform from the fixed time period $t\in[-1500M,-29M]$ to other continuous time periods, including even $12000M$ before merging. We found that the model is not limited to the dominant mode, but it can also be applied to higher-order modes, including 3-3, 2-1, 4-4, 5-5, 3-2 and 4-3 modes. We also applied the model to the spin-aligned case, but it required some adjustments. When we applied this model to the most complex eccentric spin precession situation, we were able to construct a more complete approximate model that converts the waveform without spin and eccentricity into the waveform with eccentricity and spin precession. This suggests a phenomenological and universal relationship between the eccentric waveform and the circular orbit waveform, which can help us better understand the relationship between eccentricity, spin and precession in binary black hole mergers.

This article is structured as follows: In Sec. \ref{sec:II}, we first introduce the numerical relativity waveform data we used, as well as some basic concepts related to gravitational waves in Sec. \ref{sec:II:A}. We then provide a detailed description of the eccentricity estimators in Sec. \ref{sec:II:B}, followed by a method to measure eccentricity from the eccentric waveform in Sec. \ref{sec:II:C}. Finally, we describe the fitting process for the dominant mode in Sec. \ref{sec:II:D}, and the extended research on higher-order modes in Sec. \ref{sec:II:E}, spin-aligned cases in Sec. \ref{sec:II:F}, and spin-precessed waveforms in Sec. \ref{sec:II:G}. In section \ref{sec:III}, we present the fitting parameters and overlap results for each case, and analyze them. In section \ref{sec:IV}, we provide conclusions and outlook. Throughout this article, we use geometric units where $G=c=1$. The component masses of the binary black holes are denoted by $m_1$ and $m_2$, and the total mass is denoted by $M$, which is set to unity for simplicity. The mass ratio $q$ is defined as $q = m_1/m_2$, where $m_1$ is greater than $m_2$, and $q$ only takes positive integers. The black hole's dimensionless spin vectors are denoted by $\vec{\chi}_i=\vec{S}_i / m_i^2$ for $i=1,2$.

\section{Method}\label{sec:II}
By incorporating new waveforms, we have expanded the range of parameters we studied, including the mass ratio, initial eccentricity, time period, and spin, allowing us to comprehensively investigate the model and its applicability. In this article, we aim to provide sufficient details and add significant content to fully present the essence of the model, which differs in some aspects from that of Ref. \cite{Setyawati:2021gom}.

\subsection{Numerical relativity data}\label{sec:II:A}
There are numerous numerical relativity collaborations that have conducted extensive simulations of black hole binary mergers. However, publicly available simulations with eccentricity or high mass ratios are scarce. The data we utilized in our study are sourced from two collaborations. The first is the Simulating eXtreme Spacetimes (SXS) Collaboration, which uses a multi-domain spectral method \cite{Lindblom:2005qh,Szilagyi:2009qz,Kidder:1999fv,Scheel:2008rj,Hemberger:2012jz} with a first-order version of the generalized harmonic formulation \cite{Pretorius:2004jg,Samary:2012bw,Garfinkle:2001ni,Hemberger:2012jz} of Einstein's equations with constraint damping to evolve the initial data. The Spectral Einstein Code (SpEC) \cite{SXSBBH} is used for the simulation. The SXS catalog has published 23 sets of non-spinning eccentric waveforms with mass ratios $q\in[1,3]$ and eccentricity range $e_0\in[0,0.2]$. The second set of waveforms we used was obtained from the Rochester Institute of Technology (RIT) \cite{RITBBH}. The simulations in the RIT catalog were evolved using the LazEv code \cite{Zlochower:2005bj} implementation of the moving puncture approach \cite{Campanelli:2005dd} and the BSSNOK formalism of evolution systems \cite{Healy:2019jyf,Zlochower:2005bj,Marronetti:2007wz}. The LazEv code uses the Cactus \cite{Lousto:2007rj} /Carpet \cite{Zlochower:2012fk} /EinsteinToolkit \cite{Healy:2016lce} infrastructure. The 4th release of the RIT catalog published 824 eccentric black hole binary NR simulations, including spinning, spin-aligned, and spin-precessing cases with eccentricities ranging from 0 to 1 \cite{Healy:2022wdn}. 

\begin{figure}[htb]
\centering
\includegraphics[width=\linewidth]{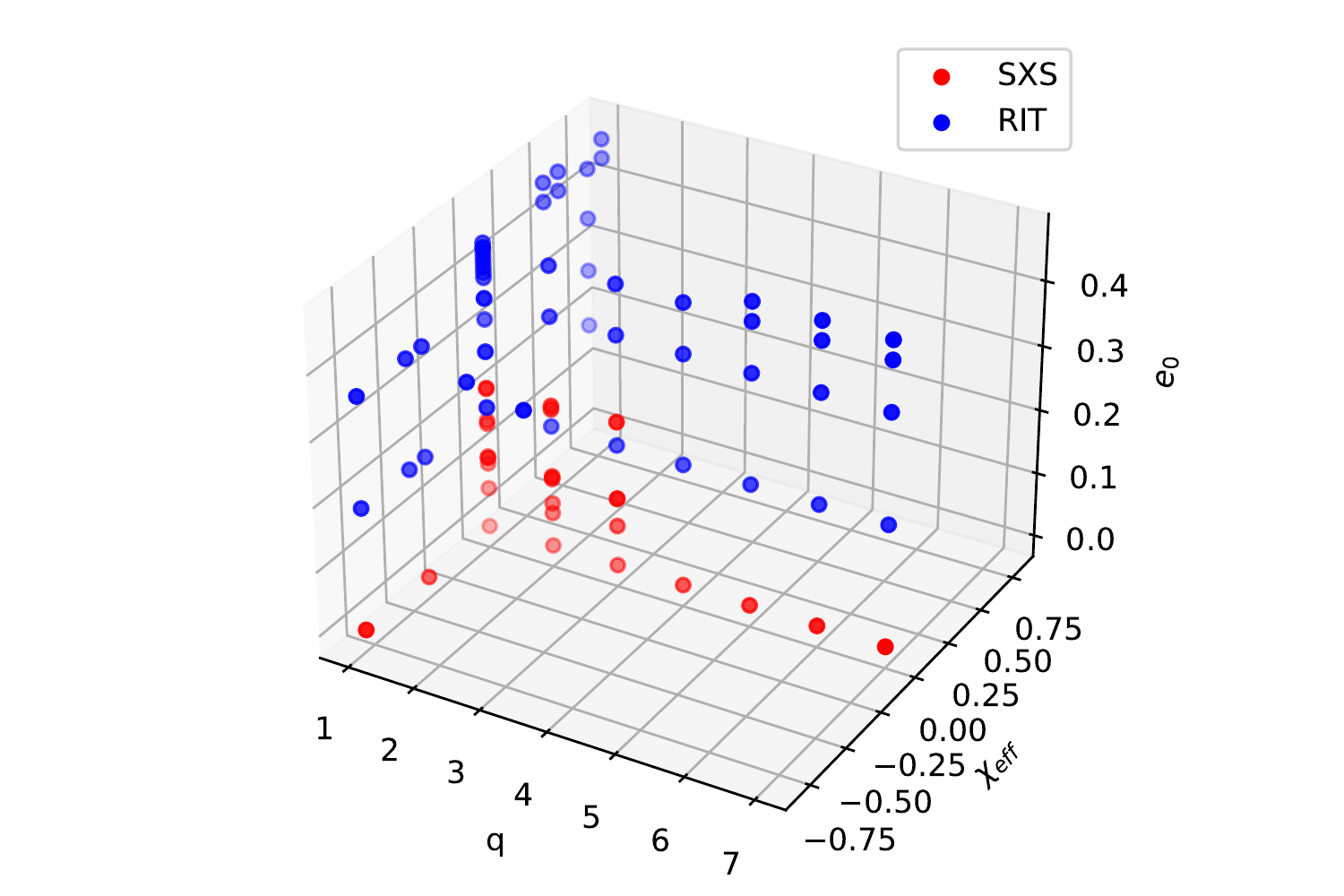}
\caption{\label{FIG.1} Waveform parameter
distribution of the data for mass ratio $q$, effective spin $\chi_{\rm eff}$ and initial eccentricity $e_0$ coming from SXS and RIT catalog.}
\end{figure}

In numerical relativity, waveforms are obtained by computing the Newman-Penrose scalar $\Psi_4$ at a finite radius and then extrapolating to null infinity. $\Psi_4$ can be expanded by the spin-weighted spherical harmonic function ${ }_{-2}Y_{\ell, m}(\theta, \phi)$ with spin weight $s=-2$ as
\begin{equation}
r\Psi_4=\sum_{\ell, m}{r\Psi_4}_{(\ell m)}{ }_{-2} Y_{\ell,m}(\theta, \phi),
\end{equation}
where $r$ is the extraction radius. The gravitational wave strain $h$ can also be expressed as
\begin{equation}
rh=r\left(h_{+}-i h_{\times}\right)=\sum_{\ell, m}rh_{\ell m}{ }_{-2} Y_{\ell, m}(\theta, \phi),
\end{equation}
where $h_{+}$ and $h_{\times}$ represent the two polarizations of gravitational waves, respectively. As $r$ goes to infinity, there is a relationship between them, given by
\begin{equation}\label{eq:3}
\Psi_{4}(t)=\frac{\partial^{2}}{\partial t^{2}}h(t).
\end{equation}
Therefore, we can obtain $h$ from $\Psi_4$ using Eq. (\ref{eq:3}). We can download all the higher harmonics modes $rh_{lm}$ and $r\Psi_{4(lm)}$ from the SXS and RIT catalog databases, which have been normalized with $r$ set to unity for simplicity. We decompose $\Psi_{4}$ and $h$ into a combination of amplitude and phase as follows:
\begin{equation}\label{eq:4}
\Psi_{4(lm)}=A_{lm}(t) \exp \left[-i \varphi_{lm}(t)\right],
\end{equation}
\begin{equation}\label{eq:5}
h_{lm}=\mathcal{A}_{lm}(t) \exp \left[-i \Phi_{lm}(t)\right],
\end{equation}
and the amplitude and frequency of $h_{lm}$ can be obtained using the following equations:
\begin{equation}\label{eq:6}
\mathcal{A}_{lm}=|h_{lm}|,
\end{equation}
\begin{equation}\label{eq:7}
\omega_{lm}=\frac{d\Phi_{lm}}{dt}.
\end{equation}
We define the effective spin in the $z$ direction, which is aligned with the direction of the orbital angular momentum $L$, as
\begin{equation}\label{eq:8}
\chi_{\mathrm{eff}}=\frac{m_1 \chi_{1,z}+m_2 \chi_{2, z}} {m_1+m_2},
\end{equation}
where $\chi_{1,z}$ and $\chi_{2, z}$ are the dimensionless spins in the $z$ direction for the two black holes.
We obtained the gravitational wave strain $rh$ from both the SXS and RIT catalogs instead of $r\Psi_{4}$. However, our analysis in Sec. \ref{sec:II:B} shows that the model is applicable to both. During the waveform processing, we removed the first 300 and 100  of the SXS and RIT waveforms, respectively, to eliminate junk radiation. The rest of the waveform processing followed Ref. \cite{Setyawati:2021gom}. Taking the 2-2 mode waveform as an example, we first obtained the amplitude and frequency using Eq. (\ref{eq:6}) and (\ref{eq:7}), found the maximum amplitude value $\mathcal{A}$, and set that moment as $t=0$. We then aligned all amplitudes and frequencies with the same mass ratio at $t=0$. Next, we chose an initial time $t_0$, and the research time period was $[t_0,0]$. However, we needed to select a long enough waveform so that, after removing the junk radiation, the waveform was still longer than the research range of $[t_0,0]$. The waveforms used and their associated parameters in the RIT and SXS catalogs are listed in TABLE \ref{tab:I} and TABLE \ref{tab:II} in Appendix \ref{App:A}. As the SXS waveform catalog does not provide initial eccentricity, we measured the eccentricity of the waveform at the first 300 using the method in Sec. \ref{sec:II:C}. For the RIT waveform, we directly used the reference eccentricity given in the catalog. However, the initial eccentricity $e_0$ only indicates the eccentricity at the initial moment of the waveform, representing the approximate research range of the eccentricity in this article. For subsequent studies, each research period $[t_0,0]$ had a corresponding initial eccentricity $e_0=e(t_0)$. The distributions of waveform parameters for mass ratio $q$, effective spin $\chi_{\mathrm{eff}}$, and initial eccentricity $e_0$ are shown in FIG. \ref{FIG.1}. We only show the range of initial eccentricity $e_0\in[0,0.45]$ as the model depends on the length of the waveform, so many shorter waveforms were excluded. Relevant analysis can be found in Sec. \ref{sec:II:C}.
\subsection{Eccentricity Estimator}\label{sec:II:B}
It is necessary to introduce eccentricity estimator in detail. According to Ref. \cite{Mroue:2010re}, eccentricity estimator is derived from  Newton's formula containing distance, orbital phase or orbital frequency that can be used to estimate eccentricity. In the Ref. \cite{Mroue:2010re}, an eccentricity estimator is defined as an oscillatory  function
\begin{equation}\label{eq:9}
e_{X}(t)= e\cos\left(\Omega{t}+\phi\right),
\end{equation}
where $X$ represents the object used to measure eccentricity, and $e$, $\Omega$ and $\phi$ is eccentricity, frequency and phase respectively. It can be promoted that $X$ can be separation, frequency, amplitude, phase, or derivative of frequency from orbital dynamics or waveform. Based on separation
\begin{equation}
d(t)=d_0\left[1+e\cos \left(\Omega t+\phi_0\right)\right]+O\left(e^2\right),
\end{equation}
where $d_0$ and $\phi_0$ are the average distance and initial phase in Newtonian orbit. We get separation eccentricity estimator:
\begin{equation}\label{eq:11}
e_{d}(t)=\left(1\right)\left(\frac{d(t)-\Bar{d}(t)}{\Bar{d}(t)}\right),
\end{equation}
where $\Bar{d}(t)$ represents the secular average to $d(t)$, and $\Bar{d}(t)$ equals $d_0$ in Newtonian gravity, and $1$ is its coefficient. Based on orbital phase:
\begin{equation}
\Phi(t)=\Phi_0+\Omega_0 t+2 e \sin \left(\Omega t\right)+O\left(e^2\right),
\end{equation}
where $\Phi_0$ and $\Omega_0$ are phase offset and average frequency. We get orbital phase eccentricity estimator:
\begin{equation}\label{eq:13}
e_\Phi(t)= \left(\frac{1}{2}\right)\left(\Phi(t)-\Bar{\Phi}(t)-\Phi_0\right),
\end{equation}
where $\Omega_0 t$ is replaced by $\Bar{\Phi}(t)$ which is the secular average to $\Phi(t)$. Taking the first time derivative with respect to the phase, we get orbital frequency eccentricity estimator:
\begin{equation}\label{eq:14}
e_\Omega(t)= \left(\frac{1}{2}\right)\left(\frac{\Omega(t)-\Bar{\Omega}(t)}{\Bar{\Omega}(t)}\right),
\end{equation}
where $\Omega_0$ is replaced by $\Bar{\Omega}(t)$ which is the secular average to $\Omega(t)$.
We can not only get the eccentricity estimator based on the orbital dynamical quantities, but we can get it from the waveform. According to Ref. \cite{Healy:2017zqj}, based on the frequency, amplitude and phase of the Weyl scalar $\Psi_4$, we can also define the associated eccentricity estimators. In the Ref. \cite{Healy:2017zqj}, according to Eq. (\ref{eq:4}), Weyl scalar $\Psi_4$ of the 2-2 mode and its frequency can be expressed as
\begin{equation}
\Psi_{4(22)}=A_{22}(t) \exp \left[i\varphi_{22}(t)\right]
\end{equation}
\begin{equation}
\varpi_{22}={\mathrm{d}}\varphi_{22} /{\mathrm{d}}t,
\end{equation}
where
\begin{equation}
\begin{aligned}
A_{22}(t) &=K_1\left(1+\frac{39}{8} e \cos \Omega t\right)+\mathcal{O}\left(e^2\right) \\
\varphi_{22}(t) &=-2 \Omega t-\frac{21}{4} e \sin \Omega t+\mathcal{O}\left(e^2\right)\\\varpi_{22}(t) &=-2 \Omega\left(1+\frac{21}{8}e \cos \Omega t\right)+\mathcal{O}\left(e^2\right),
\end{aligned}
\end{equation}
which are first-order approximation of the eccentricity, where $K_1$ is a constant.
The associated eccentricity estimators are:
\begin{equation}\label{eq:18}
\begin{aligned}
e_{A_{22}}(t)&=\left(\frac{8}{39}\right) \left(\frac{A_{22}(t)-\Bar{A}_{22 }(t)}{A_{22}(t)}\right)\\
e_{\varphi_{22}}(t)&=\left(\frac{4}{21}\right) \left[\varphi_{22}(t)-\Bar{\varphi}_{22 }(t)+\varphi_{220}\right]\\
e_{\varpi_{22}}(t)&=\left(\frac{8}{21}\right) \left(\frac{\varpi_{22}(t)-\Bar{\varpi}_{22 }(t)}{\varpi_{22 }(t)}\right),
\end{aligned}
\end{equation}
where $\varphi_{220}$ represents a phase offset to $\varphi_{22}$.
According to Eq. (\ref{eq:5}), the 2-2 mode of gravitational waves strain $h$ and its frequency can be expressed as
\begin{equation}\label{eq:19}
h_{22}=\mathcal{A}_{22}(t) e^{i \Phi_{22}(t)}
\end{equation}
\begin{equation}\label{eq:20}
\omega_{22}={\mathrm{d}}\Phi_{22} / {\mathrm{d}} t,
\end{equation}
where
\begin{equation}
\begin{aligned}
&\mathcal{A}_{22}(t)=K_2\left(1+\frac{3}{2} e \cos \Omega t\right)+\mathcal{O}\left(e^2\right) \\
&\Phi_{22}(t)=-2 \Omega t-3 e \sin \Omega t+\mathcal{O}\left(e^2\right) \\
&\omega_{22}(t)=-2 \Omega\left(1+\frac{3}{2} e \cos \Omega t\right)+\mathcal{O}\left(e^2\right),
\end{aligned}
\end{equation}
which are first-order approximation of the eccentricity, where $K_2$ is a constant. The associated eccentricity estimators are:
\begin{equation}\label{eq:22}
\begin{aligned}
e_{\mathcal{A}_{22}}(t)&=\left(\frac{2}{3}\right) \left(\frac{\mathcal{A}_{22}(t)-\Bar{\mathcal{A}}_{22 }(t)}{\mathcal{A}_{22}(t)}\right)\\
e_{\Phi_{22}}(t)&=\left(\frac{1}{3}\right) \left[\Phi_{22}(t)-\Bar{\Phi}_{22}(t)+\Phi_{220}\right]\\
e_{\omega_{22}}(t)&=\left(\frac{2}{3}\right) \left(\frac{\omega_{22}(t)-\Bar{\omega}_{22 }(t)}{\omega_{22}(t)}\right).
\end{aligned}
\end{equation}
 We can see forms are different for Eqs. (\ref{eq:11}), (\ref{eq:13}), (\ref{eq:14}), (\ref{eq:18}) and  (\ref{eq:22}). Ref. \cite{Mroue:2010re} puts the average on the denominator, but ref. \cite{Healy:2017zqj} not. What they express is the difference between a certain quantity of the eccentric waveform and its average value, and the meaning is the same. In fact, as we will see later, no matter it is based on the orbital dynamical quantity e.g. orbital distance, phase and frequency, or the gravitational waveform $h$ and $\Psi_4$, whether it is based on the 2-2 mode or the high-harmonics-order modes, it turns out that we can obtain the associated eccentricity estimator, and they can be summed up in the form:
\begin{equation}\label{eq:23}
e_{X_{1}}(t)=\left(k_1\right) \left(\frac{{X_{1}}(t)-{\Bar{X}_{1}}(t)}{{\Bar{X}_{1}}(t)}\right)
\end{equation}
\begin{equation}\label{eq:24}
e_{X_{2}}(t)=\left(k_2\right)\left[{X_{2}}(t)-{\Bar{X}_{2}}(t)+X_{20}\right],
\end{equation}
where $X_1$ represents the quantity relative to orbital distance, frequency and amplitude or frequency and amplitude of waveform, $X_2$ represents the quantity relative to phase, and $X_{20}$ is a phase offset, and $k_1$, $k_2$ can be some constant. Then we will find that $\Bar{X}$ is the quantity $X_c$ come from its corresponding circular orbit waveform which has the same time length, mass ratio and spin as the eccentric waveform.
We need to emphasize that there is a difference between quantities relative to waveform and orbital dynamics, which is reflected in the different constant coefficients of the eccentricity estimator derived as we can see in Eqs. (\ref{eq:11}), (\ref{eq:13}), (\ref{eq:14}), (\ref{eq:18}) and  (\ref{eq:22}). However, the difference in the constant coefficients $k_1$ only affects the magnitude of the Eq. (\ref{eq:23}), not the overall morphological behavior of the eccentricity estimator. So the phenomenological fitting model which we will introduce later is applicable for them.

\subsection{Measuring eccentricity of waveform}\label{sec:II:C}
\begin{figure}[htb]
\centering
\includegraphics[width=\linewidth]{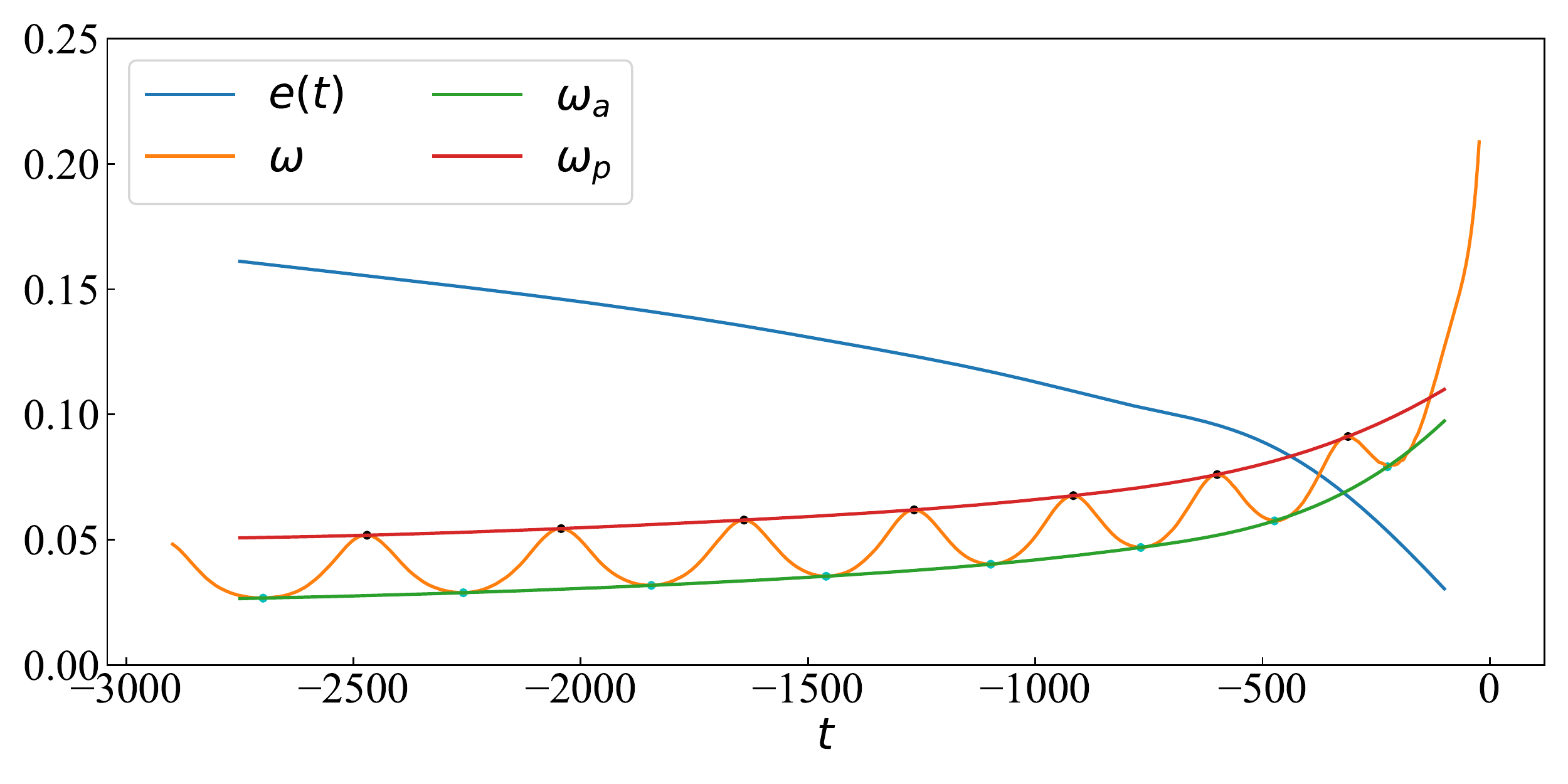}
\caption{\label{FIG:2}Time evolution of waveform frequency $\omega$, apastron $\omega_a$, periastron $\omega_p$, and eccentricity $e(t)$ for numerical simulation SXS:BBH:1360.}
\end{figure}

\begin{figure}[htb]
\centering
\includegraphics[width=\linewidth]{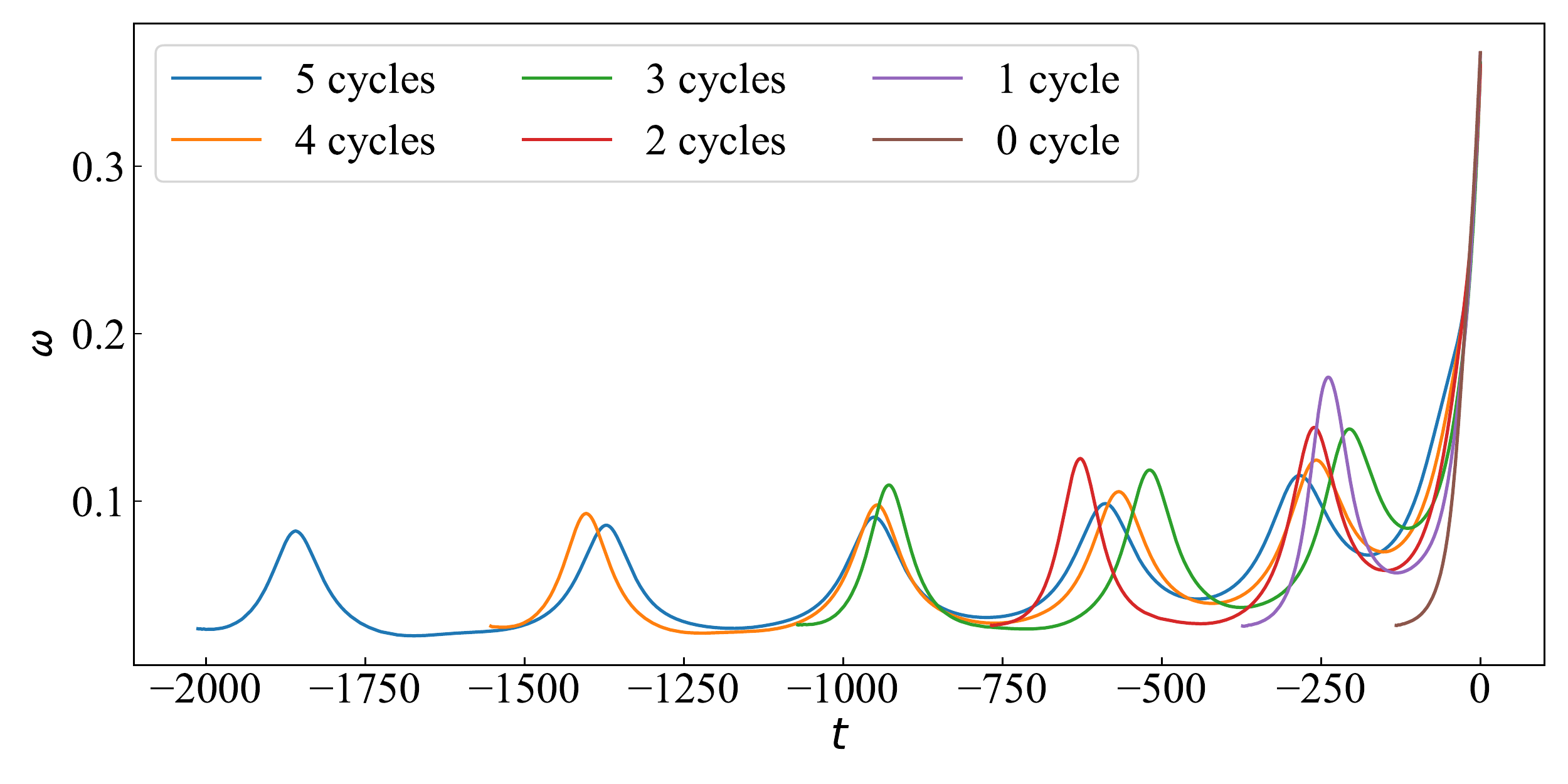}
\caption{\label{FIG:3}Frequency of waveforms which have 5 cycles, 4 cycles, 3 cycles, 2 cycles, 1 cycle, 0 cycle. For 0 cycle, 1 cycle, 2 cycles and 3 cycles, we cannot obtain the eccentricity of the waveforms, and for 4 cycles and 5 cycles, the error of the obtained eccentricity is large.}
\end{figure}

There are many definitions of eccentricity, and many ways to measure eccentricity \cite{Mroue:2010re,Healy:2017zqj,Ireland:2019tao}. Each of them have their own scope of application. In general, estimating eccentricity is mainly from the perspectives of waveform and orbital dynamics. The eccentricity estimators Eq. (\ref{eq:11}), Eq. (\ref{eq:13}), Eq. (\ref{eq:14}) to estimate the eccentricity is based on some orbital dynamical quantities. The eccentricity estimators Eq. (\ref{eq:19}), Eq. (\ref{eq:23}) to estimate the eccentricity is based on some quantities of waveform. In the Ref. \cite{Healy:2022wdn}, RIT catalog estimates the eccentricity by the dynamical coordinate distance $d$:
\begin{equation}
e_d=d^2 \ddot{d}/M.
\end{equation}
where $\ddot{d}$ represents the second derivative with respect to time. We can obtain the eccentricity at each moment by interpolating the amplitude of the eccentricity estimator in Eq. (\ref{eq:9}). But the larger the eccentricity, the more the eccentricity estimator deviates from the behavior of the \textit{cosine} or \textit{sine} function, so that it is not suitable for measuring moderate or high eccentricity. When we study high eccentricity, we have to introduce a new method capable of measuring high eccentricity according to the Ref. \cite{Ramos-Buades:2018azo}:
\begin{equation}
e_\Omega(t)=\frac{\Omega_p^{1 / 2}-\Omega_a^{1 / 2}}{\Omega_p^{1 / 2}+\Omega_a^{1 / 2}},
\end{equation}
where $\Omega$ is orbital frequency, and $\Omega_a$ and $\Omega_p$ are orbital frequency at apastron $\Omega_a$ and periastron $\Omega_p$, respectively. Although it is orbital frequency $\Omega$ here, it can apply to frequency $\omega$ of waveforms. Our approach is similar to Ref. \cite{Ramos-Buades:2019uvh}. First We use the python function \textit{find\_peaks} to find the apastron and periastron of the waveform. We then measure the eccentricity of the waveform at the apastron and periastron. Next we use the python function \textit{cubic\_spline} interpolation to obtain a continuous evolution of the eccentricity. FIG. \ref{FIG:2} shows the evolution of frequency $\omega$, apastron $\omega_a$, periastron $\omega_p$ and eccentricity $e(t)$ of the waveform SXS:BBH:1360. According to error propagation formula:
\begin{equation}
\delta e_\omega=\frac{\delta \omega}{\left(\omega_a^{1 / 2}+\omega_p^{1 / 2}\right)^2}\left[\frac{\omega_a^{1 / 2}}{\omega_p^{1 / 2}}+\frac{\omega_p^{1 / 2}}{\omega_a^{1 / 2}}\right],
\end{equation}
conservatively estimated error of frequency in the Ref. \cite{Ramos-Buades:2019uvh} is about $\delta \omega_a = \delta \omega_p=\delta \omega=0.0001$ caused by the different resolutions of the numerical simulations. The statistical error introduced by it in the measurement of eccentricity is $\delta e_\omega \approx 0.001$, which is not the main error. The main error is caused by interpolation when we use the method in FIG. \ref{FIG:2} to measure eccentricity, but it is not easy to quantify it. When the initial eccentricity is about 0.4, the interpolation error can be as high as 0.1. The fewer points that can be interpolated, the higher the error. For the research in this paper, if a waveform has more than 6 cycles and the eccentricity is lower than 0.3, the measurement error for eccentricity is not very large, but as the eccentricity increases, the measurement error becomes larger and larger. We pick frequencies of some representative waveforms which have the same initial orbital distance and large initial eccentricity, so that their number of cycles is very few. In FIG. \ref{FIG:3}, we show the frequency of waveforms RIT:eBBH:1289, RIT:eBBH:1290, RIT:eBBH:1294, RIT:eBBH:1299, RIT:eBBH:1305, and RIT:eBBH:1310 separately, which have 5 cycles, 4 cycles, 3 cycles, 2 cycles, 1 cycle, 0 cycle. For a waveform, when initial eccentricity $e_0$ is small or initial distance between the BBH is large, we can get more cycles, such as SXS:BBH:1360 in the FIG. \ref{FIG:2}. On the contrary, we can only get a few cycles. If initial eccentricity $e_0$ of the waveform is very large and close to 0.55,and initial distance is small, BBH will merge directly without any cycle such as \textit{0 cycle} in the FIG. \ref{FIG:3}. The way we measure eccentricity is by interpolating with \textit{cubic\_splines}, which requires at least four points to be interpolated. At the same time, fitting of the eccentricity estimator in Sec. \ref{sec:II:D} also requires enough waveform cycles to get enough information about mass ratio and initial eccentricity, so that we can obtain a sufficiently accurate waveform, which also determines the length of the waveform that we used is at least $[-1500,0]$ to be reliable. Therefore, for cases such as \textit{0 cycle}, \textit{1 cycle}, \textit{2 cycles} and \textit{3 cycles} in the FIG. \ref{FIG:3}, we cannot obtain the initial eccentricity of the waveforms, and for cases such as \textit{4 cycles} and \textit{5 cycles}, the error of the obtained initial eccentricity is large. For these reasons, the waveforms we can use are limited, and the range of their initial eccentricity is limited to less than moderate initial eccentricity $e_0 \sim 0.4$, but for higher initial eccentricity till to 1, the phenomenological model will no longer apply due to limitations in length of waveforms and eccentricity measurements. So we only list waveforms used in subsequent research in the Sec. \ref{sec:II:A} and many high eccentricity waveforms are omitted in RIT catalog.

\subsection{Fitting process}\label{sec:II:D}
Due to the circularization of the gravitational radiation, eccentric waveform has the same properties as the circular orbit waveform near merger \cite{Hinder:2007qu}. We only study the waveform before merger for this article. If we want to construct the complete gravitational waveform, we need to combine the circular orbital merger and ringdown waveform. We use the same notation as Ref. \cite{Setyawati:2021gom} for eccentricity estimator equivalent to taking $k_1=\frac{1}{2}$ in the Eq. (\ref{eq:23}):
\begin{equation}\label{eq:28}
e_X(t)=\frac{X_{\mathrm{e}}(t)-X_{\mathrm{c}}(t)}{2 X_{\mathrm{c}}(t)},
\end{equation}
where $X$ represents the amplitude $\mathcal{A}$ or frequency $\omega$, and the subscripts $e$ and $c$ represent the eccentric and the circular orbital waveform, respectively. We use Eq. (3) in Ref.\cite{Setyawati:2021gom} to fit functional relationship between eccentricity estimator and circular orbit frequency or amplitude with same mass ratio in a specific time period prior to merger. Due to the monotonic correspondence between frequency, amplitude and time for circular orbital waveform, there is an assumption implicit in the Ref. \cite{Setyawati:2021gom} for the fitting model:
\begin{equation}\label{eq:29}
\begin{aligned}
e_X\left(t\right)&=e_X\left(t(X_c)\right)=\frac{X_{\mathrm{e}}\left(t(X_c)\right)-X_{\mathrm{c}}}{2 X_{\mathrm{c}}}
\\&=\frac{X_{\mathrm{e}}\left(X_c\right)-X_{\mathrm{c}}}{2 X_{\mathrm{c}}}=A e^{B X_c^\kappa} \sin \left(f X_c^\kappa+\varphi\right),
\end{aligned}
\end{equation}
where $A$, $B$, $f$, $\varphi$ and $\kappa$ are fitting parameters.
We explicitly write the functional relationship between the quantities, because it is not obvious that there is a functional relationship between the eccentric quantities $X_e$ and the circular quantities $X_c$. Instead of taking $\kappa$ as a fixed value -59/24 for amplitude or -83/24 for frequency as in the Ref. \cite{Setyawati:2021gom}, we regard $\kappa$ as a parameter which will allow us to generalize the fitting to other time periods, mass ratio and higher-order modes. It is some coincident that the Ref. \cite{Setyawati:2021gom} uses $\kappa$ as a fixed constant to fit and obtain good results, because they only use the 2-2 mode with time period $t\in[-1500,-29]$ and mass ratio $q\in[1,3]$. According to the analysis in Sec. \ref{sec:II:B}, the constant $k_1=\frac{1}{2}$ in the formula can be any other constant, which does not affect the fitting result of the Eq. (\ref{eq:29}) to the waveform. We still use it for convenience. Unlike any other fitting model to get a local fitting to waveform, the model is a global fitting to the waveform, which reflects  global nature of waveform. The fitting uses the python function \textit{op.curve\_fit} that uses the non-linear least square fitting method to obtain the best fit.

We can fit the quantities such as amplitude $\mathcal{A}$ and frequency $\omega$ of waveform well through Eq. (\ref{eq:29}), and then look for  relationships between initial eccentricity $e_0$, mass ratio $q$ and fitting parameters $A$, $B$, $f$, $\varphi$ and $\kappa$ in a specific time period. When the data of the waveforms is not so much, it is difficult for us to discover the relationships between them. After adding waveforms from RIT catalog, the data is more and the relationship between the parameters is more obvious. Obtained the relationship between fitting parameters and waveform parameters, we can cover the entire parameter range by interpolation or polynomial fitting. Then by inverting Eq. (\ref{eq:29}), we get the amplitude or frequency of the eccentric waveform we want as follows:
\begin{equation}\label{eq:30}
\begin{aligned}
X_{e}(t)&=X_{e}(t(X_{c}))=X_{e}(X_{c})\\
&=2X_{c}e_X\left(t(X_c)\right)+X_{c}
\\&=2X_{c}A e^{B X_c^\kappa} \sin \left(f X_c^\kappa+\varphi\right)+X_{c}.
\end{aligned}
\end{equation}

In order to improve the fitting effect, we may set $\kappa$ to different $\kappa_1$ and $\kappa_2$:
\begin{equation}
e_X\left(t\right)=A e^{B X_c^{\kappa_1} }\sin \left(f X_c^{\kappa_2}+\varphi\right).
\end{equation}
But the results of the fitting show that it does not work, because

(i) Larger fitting parameter space makes it difficult to find relationships between fitting parameters and waveform parameters. The error is larger when we obtain eccentric waveforms through the Eq. (\ref{eq:30}).

(ii) Existence of  parameter $\kappa_1$ makes $e^{B X_c^{\kappa_1}}\approx1$ unstable, thus destroying the magnitude relationship $A\approx A e^{B X_c ^{\kappa}}$, thereby destroying the proportional relationship between $A$ and eccentricity $e_0$, and introducing a larger error. It is very important that eccentricity $e_0$ is proportional to $A$, because it makes the fitting more accurate. So we must maintain the relationship.

\begin{figure*}[htbp!]
\centering
\includegraphics[width=15cm,height=5cm]{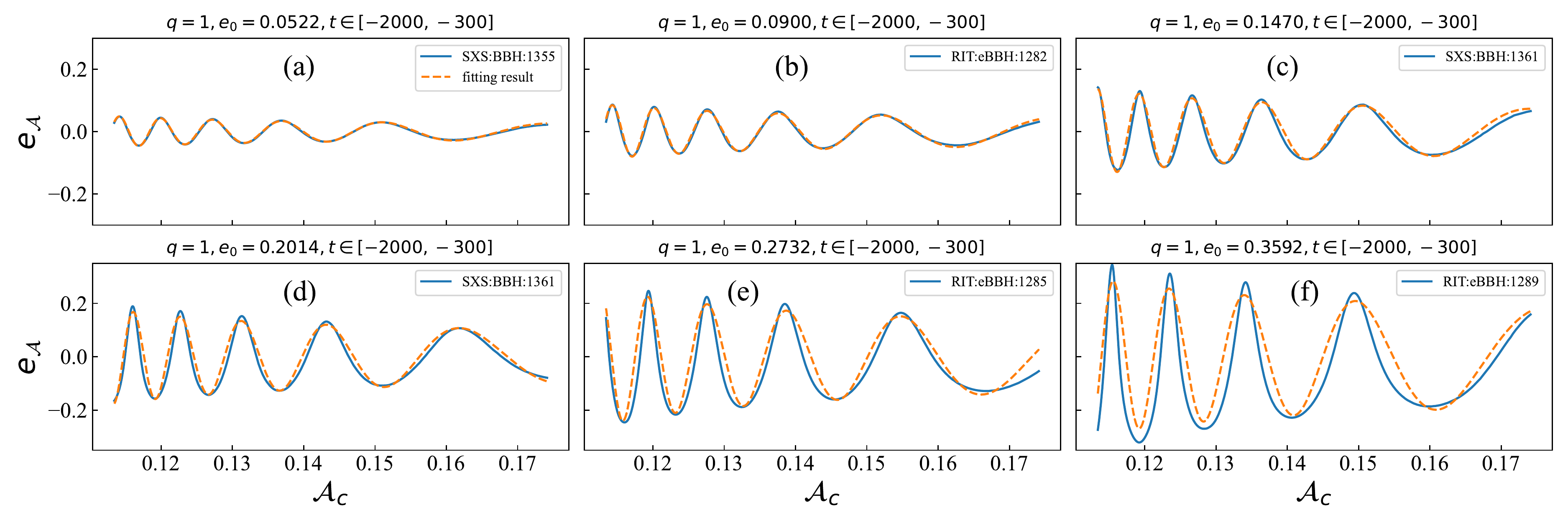}
\caption{\label{FIG:4}Fitting results for amplitude $\mathcal{A}$ of waveforms of 2-2 mode of the same $q=1$, time range $t\in[-2000,-300]$ but different initial eccentricities.}
\end {figure*}

Unlike the Ref. \cite{Setyawati:2021gom}, we do not want to construct a complete inspiral-merger-ringdown waveform, but study the waveform of 300 prior to merger. Near merging, eccentricity is very small and negligible.  Fitting results show that the part close to merging cannot be well fitted, and forced fitting will only bring errors. We give the fitting results of several different cases as follows:

(i) The same mass ratio $q=1$, the same time period $t\in[-2000,-300]$, and different initial eccentricity $e_0$ (see FIG. \ref{FIG:4}).

\begin{figure*}[htbp!]
\centering
\includegraphics[width=15cm,height=5cm]{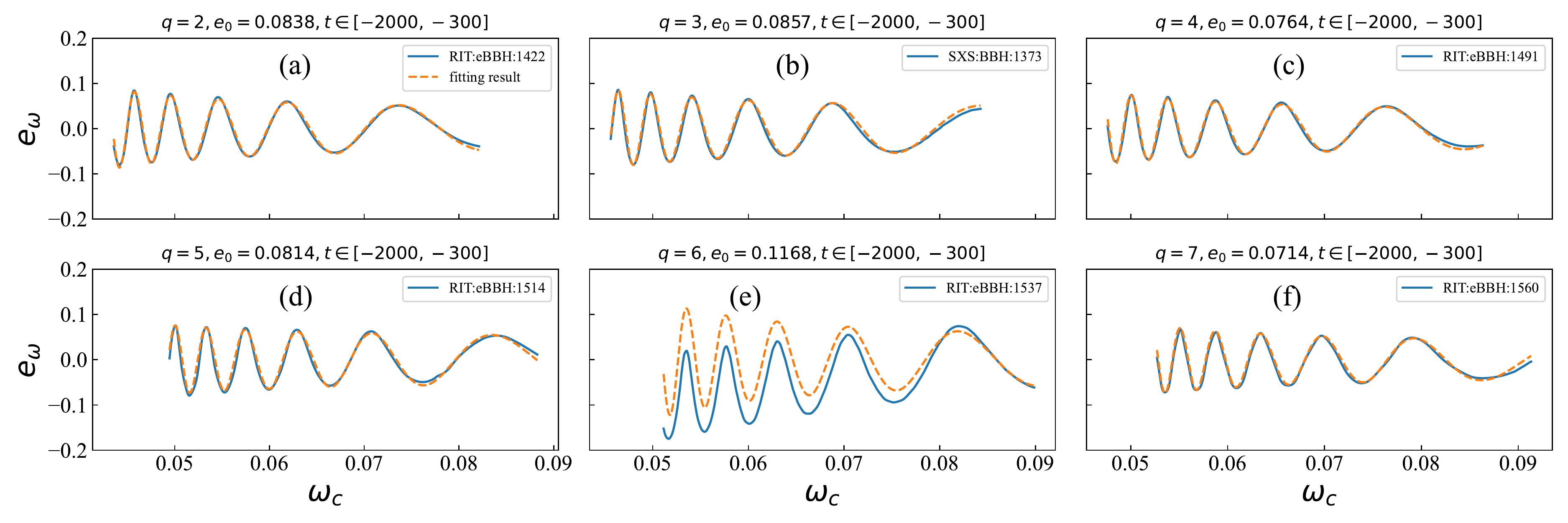}
\caption{\label{FIG:5}Fitting results for frequency $\omega$ of waveforms of 2-2 mode of the similar initial eccentricity, time range $t\in[-2000,-300]$ but different mass ratio $q\in[2,7]$.}
\end {figure*}

(ii) Different mass ratios $q\in[1,7]$, the same time period $t\in[-2000,-300]$, and similar initial eccentricity $e_0$ (see FIG. \ref{FIG:4} (b) and FIG. \ref{FIG:5})

\begin{figure*}[htbp!]
\centering
\includegraphics[width=15cm,height=5cm]{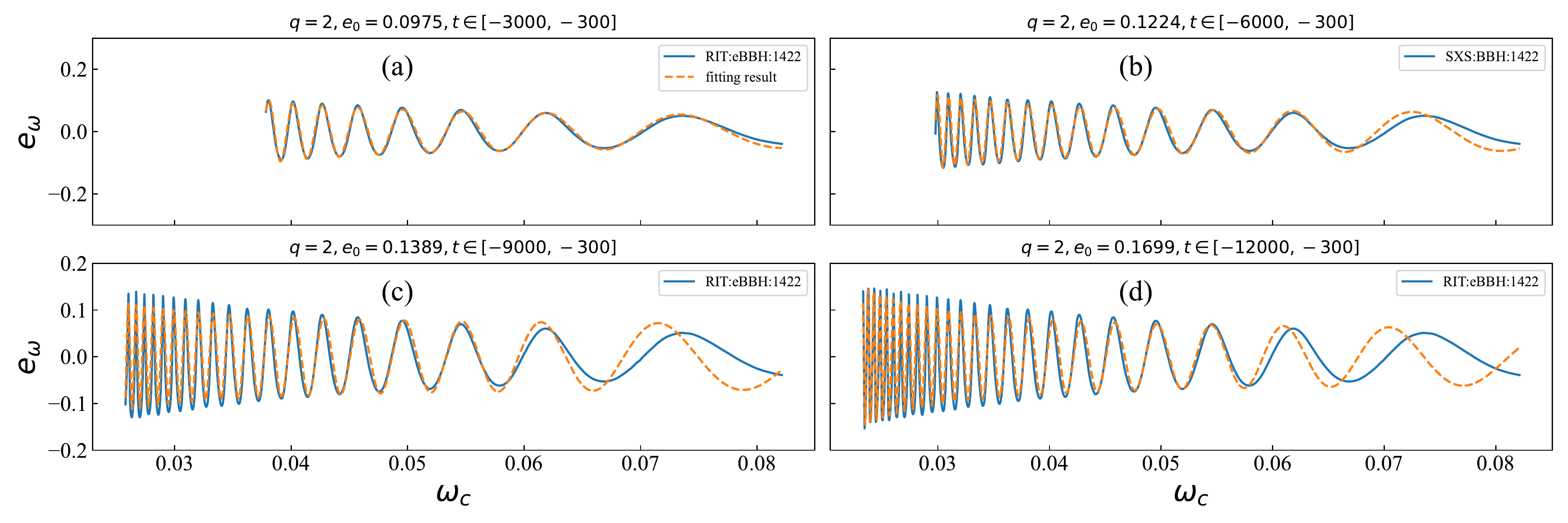} 
\caption{\label{FIG:6}Fitting results for frequency $\omega$ of waveforms of 2-2 mode of  RIT:eBBH:1422 for different time range.}
\end {figure*}

(iii) The same mass ratio $q=1$, different time periods, and different initial eccentricity $e_0$ for numerical simulation RIT:eBBH:1422 (see FIG. \ref{FIG:6}).

From fitting results of each case, we can draw some conclusions:

(i) From FIG. \ref{FIG:4}, we find the larger the initial eccentricity, the worse the fitting effect. When the initial eccentricity is very small, for $e_0=0.0522$, we can almost achieve perfect fitting, but when $e_0=0.3592$, we cannot achieve good fitting. For the case with larger $e_0$, it is not easy to obtain a waveform in time period  $t\in[-2000,-300]$, because it requires BBH to have a larger initial separation.

(ii) From FIG. \ref{FIG:4} (b) and FIG. \ref{FIG:5}, we can see the fitting model can be applied to mass ratios $q\in[1,7]$. We find that a poor fitting result is obtained in FIG. \ref{FIG:5} (e), because there are some problems with the numerical simulation RIT:eBBH:1357 itself, not with the fitting model.

(iii) From FIG. \ref{FIG:6}, we find the fitting model can be applied to a very long time period. Even for $t\in[-12000,-300]$, the morphology of the waveform can be roughly grasped.

For FIG.\ref{FIG:4}, in the following Sec. \ref{sec:III} we will discuss how good the results are for different eccentricity fits, and how much initial eccentricity can be applied. For FIG.\ref{FIG:5}, we only list fitting results of mass ratio 1-7 waveforms here, due to the limitation of waveform data. However, based on the origin of the eccentricity estimator using Newton's approximation to estimate eccentricity, we believe that this model can be extended to other situations beyond mass ratio 7. For FIG.\ref{FIG:6}, we found that for very long waveforms, this phenomenological model can roughly capture its morphology.  However, the longer the waveform, the worse the fit for the tail of the waveform. We found it increasingly difficult to obtain optimal fit parameters as longer waveforms were used. The fit to the waveform will also get worse and worse, as discussed in subsequent Sec. \ref{sec:III}. So the fit is finite and cannot extend the time period to negative infinity. In fact, $3000M$ is already a long enough waveform for numerical relativity.
\subsection{Extend to higher-order modes}\label{sec:II:E}
\begin{figure}[htb!]
\centering
\includegraphics[width=5cm,height=5cm]{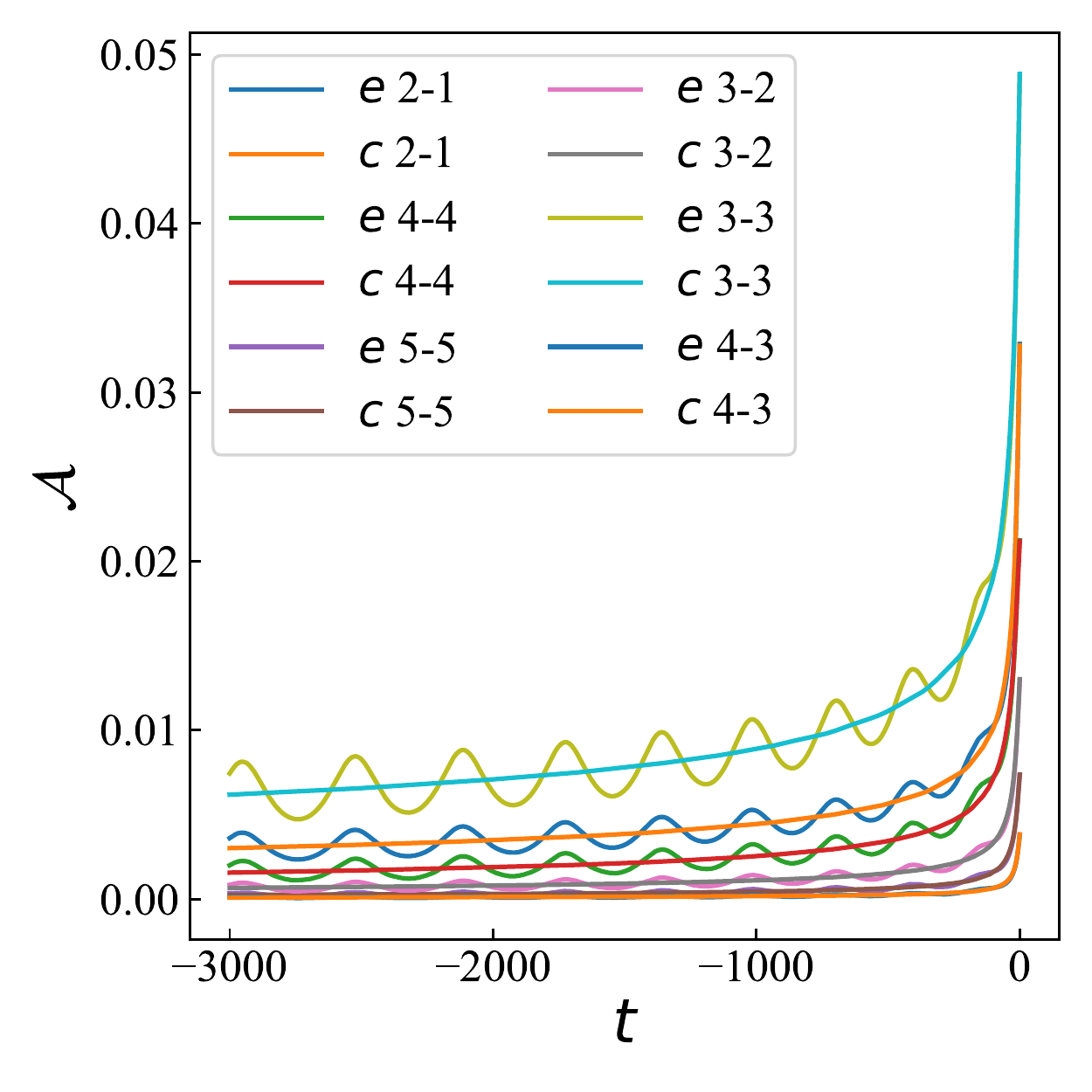}
\caption{\label{FIG:7}Amplitude $\mathcal{A}$ of high-order modes including 3-3, 2-1, 4-4, 5-5, 3-2, 4-3 modes of waveform SXS:BBH:1368 expressed as $e$ and its corresponding circular orbit waveform SXS:BBH:1165 expressed as $c$.}
\end{figure}

\begin{figure*}[!htbp]
\centering
\includegraphics[width=15cm,height=5cm]{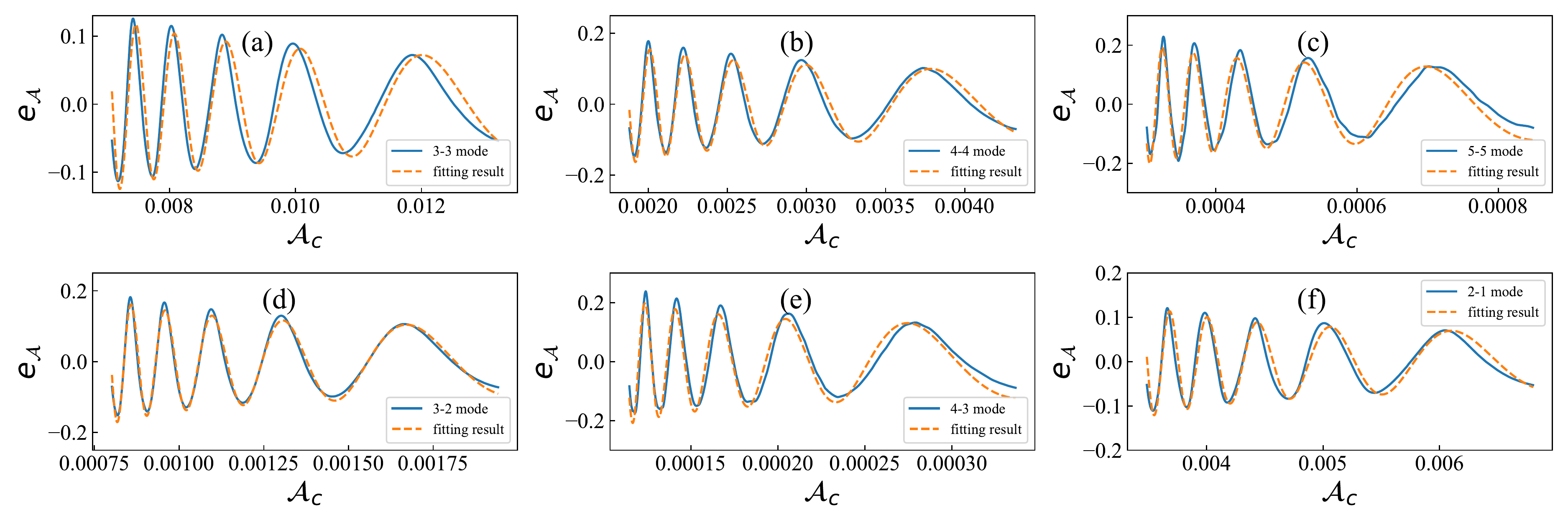}
\caption{\label{FIG:8}Fitting results for amplitude $\mathcal{A}$ of 3-3,4-4,5-5,3-2,4-3,2-1 modes of waveform SXS:BBH:1368 for the same time period $t\in[-2000,-300]$ and initial eccentricity $e_0=0.0929$. }
\end{figure*}

In fact, this model can be applied not only to 2-2 mode, but also to higher-order modes. Exactly the same as 2-2 mode, we must maintain the one-to-one correspondence between the eccentric waveform and the circular waveform.
As an example, when we study the eccentric nonspinning 2-1 mode, we have to use the associated circular nonspinning 2-1 mode. Here, we only list the 3-3, 2-1, 4-4, 5-5, 3-2, 4-3 modes of some waveforms given in the SXS catalog, because some other modes are not given in the catalog, and there are not many high-order modes of waveforms in the RIT catalog. In FIG. \ref{FIG:7}, we can see that there is a one-to-one correspondence between the  eccentric nonspinning high-order modes  and  circular ones, which makes the same fitting model applicable to them. Here, we only take numerical simulation SXS:BBH:1368 with mass ratio $q=2$, time period $t\in[-2000,-300]$ and initial eccentricity $e_0=0.0929$ as an example to give fitting result for higher-order modes. In FIG. \ref{FIG:8}, we present the fitting results of amplitude eccentricity estimator $e_{\mathcal{A}}$ for each modes, which are consistent with the low eccentricity of the 2-2 mode. In fact, we can also consider other situations, and will find the variation behavior of all the higher-order modes is exactly the same as the 2-2 mode, which means that the model can also be applied to higher-order modes with other mass ratios, other eccentricities, and other time period.
\subsection{Extend to spin-aligned}\label{sec:II:F}
Spin is an important parameter of gravitational waves, which is very necessary to contain to characterize nature of source. From eccentric numerical simulations RIT:eBBH:1282, RIT:eBBH:1740, RIT:eBBH:1763 and RIT:eBBH:1899, we can see that in these waveforms anti-aligned spin has the same effect as eccentricity to speed up BBHs merger compared to circular orbital nonspinning waveform. We discover that we can establish a relationship between eccentric spin-aligned or spin-anti-aligned waveforms and circular orbit waveforms from two perspectives with the phenomenological fit model. For the convenience of description, we now collectively refer to spin alignment or spin anti-alignment as spin alignment, which is the same for this model.

\begin{figure*}[!htbp]
\centering
\includegraphics[width=15cm,height=2.5cm]{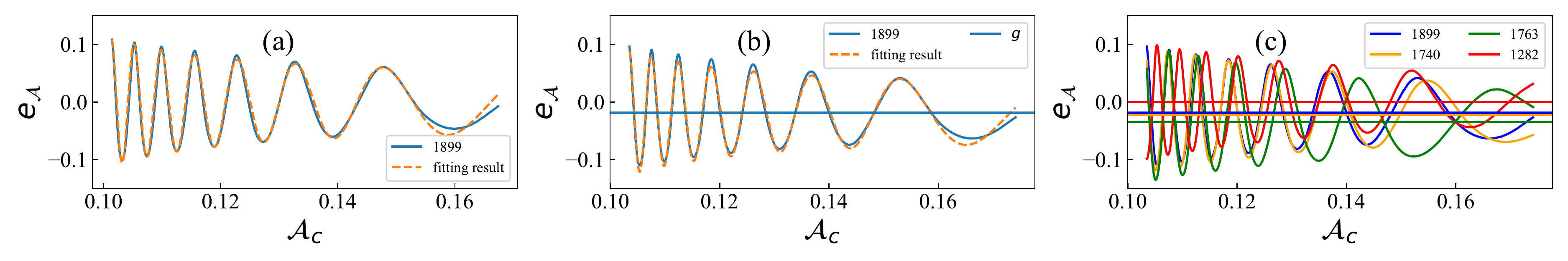}
\caption{\label{FIG:9}From left to right. Panel (a) is the fitting result of the amplitude eccentricity estimator of waveform RIT:eBBH:1899, which comes from a circular orbit waveform with the same spin as it. Panel (b) is the fitting result of the amplitude eccentricity estimator of waveform 1899, which comes from a circular orbit waveform without spin, the line $g$ shows the effect of aligned-spin on it. Panel (c) are amplitude eccentricity estimators of waveform RIT:eBBH:1282, RIT:eBBH:1740,  RIT:eBBH:1763, RIT:eBBH:1899, which come from a circular orbit waveform without spin, the corresponding colors horizontal lines $g$ shows the effect of aligned-spin on them.}
\end{figure*}

(i) As in the previous case, we can establish a relationship as Eq. (\ref{eq:29}) between eccentric spin-aligned waveforms and circular orbit spin-aligned waveforms. We must maintain the consistency of spins for eccentric waveforms and  circular waveforms. For example, if initial dimensionless spins of BBH ${\chi_z}_1=-0.5$, ${\chi_z}_2=-0.5$ for one of the waveforms, it must be the same for another. Here, we take the eccentric waveform as RIT:eBBH:1899, whose $q=1$, two dimensionless spins are ${\chi_z}_1=-0.5$, ${\chi_z}_2=-0.5$, time period is $t\in[-3000,-300]$ and initial eccentricity $e_0=0.1110$, and the circular orbit waveform as SXS:BBH:0325, which has the same mass ratio, time range and dimensionless spin. In FIG. \ref{FIG:9} (a), we present the fitting result.

\begin{figure}[!htbp]
\centering
\includegraphics[width=5cm,height=5cm]{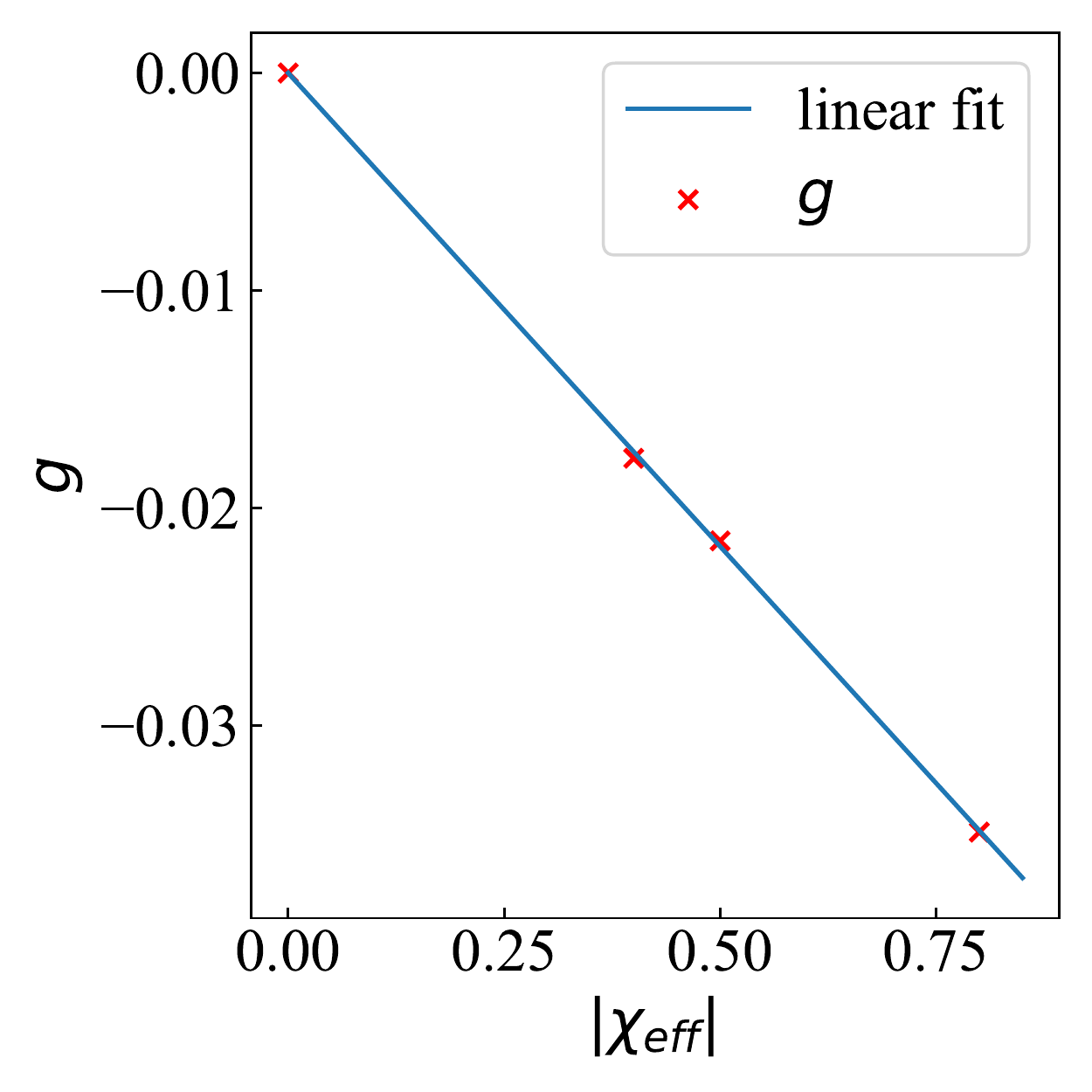}
\caption{\label{FIG:10}The red 'x's are the relationship between $g$ from the FIG. \ref{FIG:8} (c) and effective spin $\left|\chi_{\mathrm{eff}}\right|$, and the blue line represents a linear fit to them.}
\end{figure}

(ii) We try another way. We keep the waveform RIT:eBBH:1899, but we take the nonspinning circular orbital waveform SXS:BBH:0180 not the spin-aligned circular orbital SXS:BBH:0325 as seed circular orbital waveform. We find that if we want to fit the waveform, we have to introduce a shift parameter $g$ in the Eq. (\ref{eq:29}) as follows:
\begin{equation}\label{eq:32}
   e_X\left(t\right)=A e^{B X_c^\kappa} \sin \left(f X_c^\kappa+\varphi\right)+g.
\end{equation}
The parameter $g$ reflects the influence of aligned spin on the waveform and straight line $e_{\mathcal{A}}=g$ is the approximate axis of symmetry of the waveform. The phenomenological shift parameter ``$g$'' introduced here is only an approximate attempt, and its applicability depends on the result of the fitting. In FIG. \ref{FIG:9} (b), we show the fitting result and the position of the parameter $g$. We discover we can obtain a good fit for  spin-aligned waveform. We select the waveforms RIT:eBBH:1282, RIT:eBBH:1740, RIT:eBBH:1763, RIT:eBBH:1899 to further study the influence of spin on the fitting effect. We find that, aligned spin makes the entire waveform translate $g$ to the negative half axis of the $e_{\mathcal{A}}$-axis. The larger the effective spin, the greater the translation effect (see FIG. \ref{FIG:9} (c)). We discover there is a strictly proportional relationship between $g$ and the absolute value of effective spin $\chi_{\rm eff}$ of BBH(see FIG. \ref{FIG:10}).
It can be expressed as
\begin{equation}\label{eq:33}
   g = a\left| \chi_{\rm eff} \right|,
\end{equation}
where a=-0.04355, obtained by linear fitting. What we need to emphasize is that the waveforms in the FIG. \ref{FIG:9} (c) do not have the same initial eccentricity at time $t=-3000$, because both eccentricity and spin have an impact on the evolution of the waveform. If we want to obtain a counterpart of RIT:eBBH:1899 with the same eccentricity from RIT:eBBH:1282, we can use the method introduced in Sec. \ref{sec:III:A}. 

The above is just take amplitude as an example. For frequenciy we can also achieve the same result. With a simple shift parameter ``$g$", we are able to obtain a good fit to eccentric spin-aligned waveform, which is surprisingly successful. In the circular orbit limit and PN approximation, relationship between spinning waveform and nonspinning waveform cannot be described by a simple shift constant. Based on Ref. \cite{Henry:2022dzx}, we observe that at the zero eccentricity limit, the effective spins and $g$ are proportionally related to the contribution from the leading 1.5PN, 2.5PN, and 3PN orders of align-spin. However, a nonlinear relationship exists with the spin, as the quadratic term only emerges at the 3.5PN order. As a result, we conclude that this approach can achieve a certain level of accuracy. In the waveform database, there are only three sets of spin-aligned waveforms with moderate eccentricity and mass ratio of 1, which may be a loss of generality for linear fits. But combining the results for nonspinning waveform RIT:eBBH:1282, we find that they showing good linearity, and we can get the same linear result for the frequency fit, which also implies that although there are few data points, they do have such a relationship.

To summarize, for the case of spin-align waveform, we can use two kinds of seed circular orbital waveforms, one with spin and the other with no spin. The former reflects the phenomenonal model can be applied to the waveform with spin, and the latter is an approximate attempt to generate eccentric spinning waveform through nonspining circular orbital waveform. In both cases, we have obtained good fitting result in Sec. \ref{sec:III}. If we need to obtain a spin-aligned eccentricity-orbit waveform amplitude $\mathcal{A}_{e,s}$ from a circular-orbit nonspinning waveform amplitude $\mathcal{A}_c$, we can use the following equation:
\begin{equation}\label{eq:34}
\begin{aligned}
\mathcal{A}_{e, s} =2 \mathcal{A}_c\left[A e^{B \mathcal{A}_c^\kappa} \sin \left(f \mathcal{A}_c^\kappa+\varphi\right)  +a|\chi_{\text {eff }}|\right]+\mathcal{A}_c
\end{aligned},
\end{equation}
where $e$ and $s$ represent the eccentricity and align-spin, respectively. For frequency $\omega_{e,s}$ we also have the same equation as Eq.(\ref{eq:34}). In the zero eccentricity limit, Eq.(\ref{eq:34}) reduces to
\begin{equation}\label{eq:35}
\mathcal{A}_s=2 \mathcal{A}_c a\left|\chi_{\mathrm{eff}}\right|+\mathcal{A}_c,
\end{equation}
which is from PN leading orders contribution.

\subsection{Extend to spin-precession}\label{sec:II:G}
When the spin direction of BBHs is inconsistent with the direction of orbital angular momentum, it will cause orbital plane precession. Just like effective spin $\chi_{\rm eff}$, Ref. \cite{Schmidt:2014iyl} introduces an effective precession spin parameter to describe the precession effect:
\begin{equation}\label{eq:36}
\chi_p=\frac{S_p}{A_2 m_2^2},
\end{equation}
where 
\begin{equation}\label{eq:37}
\begin{aligned}
S_p & :=\frac{1}{2}\left(A_1 S_{1 \perp}+A_2 S_{2 \perp}+\left|A_1 S_{1 \perp}-A_2 S_{2 \perp}\right|\right) \\
& \equiv \max \left(A_1 S_{1 \perp}, A_2 S_{2 \perp}\right)
\end{aligned},
\end{equation}
where $S_{i \perp}$ ($i=1,2$) represents the component perpendicular to the orbital angular momentum, $A_1=2+3 q / 2$ and $A_2=2+3 /(2q)$.

\begin{figure*}[!htbp]
\centering
\includegraphics[width=15cm,height=2.5cm]{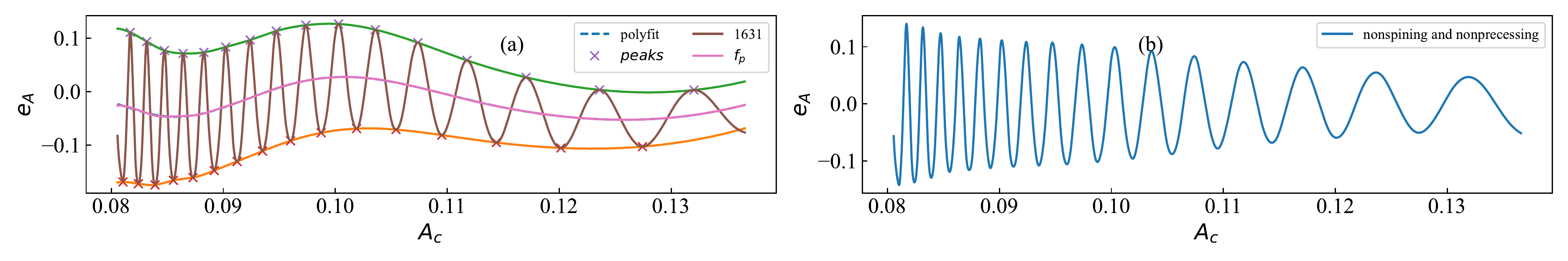}
\caption{\label{FIG:11}Figure (a) is amplitude eccentricity estimator of waveform BBH:eBBH:1631 which has been shifted $|g|=0.03$ towards the positive semi-axis of $e_{\mathcal{A}}$-axis. $f_p$ represents the precession effect from an analytical polynomial fit. Figure (b) is a non-spinning, non-precessing amplitude eccentricity estimator of waveform after subtracting the precession effect $f_p$.}
\end{figure*}

The spin-precessing eccentric waveform in RIT catalog is extremely rare. In principle, we can generate a eccentric waveform from two perspectives of spin and non-spinning circular orbital waveform as in Sec. \ref{sec:II:F}, but since the corresponding waveform of the circular orbit with similar spin is lacking, here we only study non-spinning case. We take the numerical simulation BBH:eBBH:1631 as study object which has spin $\chi_{1x}=0.7$ and $\chi_{2x}$=0.7, long time range $t\in[-12000,0]$, and low initial eccentricity $e_0=0.19$ and the other spin components are 0. 
In FIG. \ref{FIG:11} (a), we present the waveform BBH:eBBH:1631, which has been shifted $|g|=0.03$ towards the positive semi-axis of the $e_{\mathcal{A}}$-axis. We obtained this value through Eq. (\ref{eq:33}), where the effective spin of the $x$ component is $|\chi^x_{\rm eff}|=0.7$. Here, we note that prior to this work, there was no concept of an effective spin on the $x$ component, denoted as $\chi^x_{\rm eff}$, and it is different from the effective precession spin parameter $\chi_p$. We introduce this concept since $\chi^x_{\rm eff}$ plays a similar role to $\chi_{\rm eff}$, but to illustrate its physical origin, we add a superscript $x$.
According to Eq. (4.17a) in Ref. \cite{Arun:2008kb}, for the 2-2 mode, the precessing spin effect is from the 1PN order, and the align-spin effect is from the 1.5PN order. They are parameterized as the symmetric spin $\chi^x_s=(\chi_{1x}+\chi_{1x})/2$ and antisymmetric spin $\chi^x_a=(\chi_{1x}-\chi_{1x})/2$ on the $x$ component, with the $y$ and $z$ components being the same. We note that the waveform BBH:eBBH:1631 has a special configuration, leading to $\chi^x_a$ and other components being zero. Thus, $\chi^x_{\rm eff}$ and the effective spin parameter $\chi_{\rm eff}$ play a similar role.
In principle, $\chi^x_{\rm eff}$ and $\chi_{\rm eff}$ come from different PN orders, and the proportional coefficient $a$ in Eq. (\ref{eq:33}) should have different magnitudes. However, the circular-orbit nonspinning waveforms amplitude $\mathcal{A}_c$ and frequency $\omega_c$ in front of $a$ rescale the process, making the factors $a$ of both approximately equal (see Eq. (\ref{eq:35}) and Eq. (\ref{eq:42})). We note that they should not be absolutely equal due to the contribution of PN higher-order terms.
Shifting $g=0.03$ removes the effective spin effect $\chi^x_{\rm eff}$ of the waveform, leaving only the eccentricity and precession effects. 
In general, spin precession effect is very complicated. Eq. (\ref{eq:36}) quantitatively describes that the strength of the precession effect is related to the effective precession spin. But it only represents the average value of the strength of the precession effect, from which we cannot get complex modulation brought by the precession effect to the gravitational waveform. It is not easy to obtain the precessing effect of the waveform unless we have the corresponding circular orbital precession waveform. However, since the spin setting of  BBH:eBBH:1631 corresponds to a strong precession effect in which spin angular momentum and orbital angular momentum are perpendicular, its precession effect is very obvious because time scale of precession and time scale of eccentricity can be clearly distinguished. When the eccentricity of the waveform is relatively low, we can interpolate the peaks of the waveform, and obtain the midpoint of the upper and lower peaks. The reason for this operation is that the symmetry axis of the nonspinning waveform is approximately located at the midpoint of peaks of the waveform. We can accurately model the precession effect by a polynomial fit, but any other analytical method is also possible. We analytically express the precession effect as follows:
\begin{equation}\label{eq:38}
f_{p}=\sum_{i=0}^{n}a_{i}{{\mathcal{A}}_c}^i,
\end{equation}
where $a_i$ is polynomial fit coefficient. In FIG. \ref{FIG:11} (a), for the sake of accuracy, we fit the precession effect by a 10 order polynomial fit. If we express the effective spin effect $\chi^x_{\rm eff}$ as $f_{s}=g$, then we can get a nonspinning and nonprecessing amplitude eccentricity estimator by
\begin{equation}\label{eq:39}
\begin{aligned}
{e_\mathcal{A}}_{nons,nonp}={e_\mathcal{A}}_{s,p}-f_{s}-f_{p},
\end{aligned}
\end{equation}
where the subscripts $nons$ and $nonp$ stand for nonspinning and nonprecessing, respectively, and $s$ and $p$ stand for effective spin and precession.
If we want to obtain a precessing and spining amplitude eccentricity estimator, we just invert Eq. (\ref{eq:39}). In the FIG. \ref{FIG:11} (b), we subtract the precession effect and then obtain an $e_\mathcal{A}$ without effective spin or precession. This leads to a new eccentricity estimator that is exactly morphologically similar to nonspinning eccentricity estimator we encountered earlier. This basically proves that our operation of "fitting out" the precession effect is correct, and the results in Sec. \ref{sec:III:B} verify its correctness. However, what we want to emphasize is that the waveform in FIG. \ref{FIG:11} (b) are not the same as nonspinning RIT:eBBH:1282, due to the influence of spin and eccentricity on the evolution of the waveform, whose imprints have been left in it.

In summary, incorprating Eq. (\ref{eq:29}), Eq. (\ref{eq:33}) and Eq. (\ref{eq:38}), if we denote the eccentricity effects as $f_{e}$, we can obtain an eccentric spin-precessing $e_\mathcal{A}$ by gradually adding these effects to a nonspinning circular orbit waveform, which can be described as follows:
\begin{equation}\label{eq:40}
\begin{aligned}
{e_\mathcal{A}}_{e,s,p}&=\mathcal{A}_c+f_{e}+f_{s}+f_{p}\\&=A e^{B \mathcal{A}_c^\kappa} \sin \left(f \mathcal{A}_c^\kappa+\varphi\right)+a\left|\chi^x_{ \rm ef f}\right|\\&+\sum_{i=0}^{n} a_i {{\mathcal{A}}_c}^i.
\end{aligned}
\end{equation}
Then, we obtain the corresponding amplitude by the Eq. (\ref{eq:30})
\begin{equation}\label{eq:41}
\begin{aligned}
{\mathcal{A}}_{e,s,p}&=2{\mathcal{A}}_c\left(\mathcal{A}_c+f_{e}+f_{s}+f_{p}\right)+{\mathcal{A}}_c\\
&=2{\mathcal{A}}_c[A e^{B \mathcal{A}_c^\kappa} \sin \left(f \mathcal{A}_c^\kappa+\varphi\right)\\
&+a\left|\chi^x_{\rm eff}\right|
+\sum_{i=0}^{n}a_i{{\mathcal{A}}_c}^i] +{\mathcal{A}}_c.
\end{aligned}
\end{equation}
We can also get the frequency $\omega_{e,s,p}$ by the same procedure. Both of them are fundamental components of gravitational waves.

It seems miraculous that Eq. (\ref{eq:41}) can use some simple ideas and parameters to generate a spin-precession waveform of eccentricity from a nonspinning waveform of a circular orbit. But it should be noted that, as we mentioned in Subsec. \ref{sec:II:F}, for the treatment of effective spin, we adopted the leading-orders approximation, and for the precession effect, we only considered the simplest case of strong precession in special configurations. How accurate this approximation depends on the comparison of the generated waveform to the original waveform. 

Eq. (\ref{eq:41}) may also be applicable to the effective spin of the $y$ component, denoted as $\chi^y_{\rm eff}$, since $x$ and $y$ are at the same level and have the same form in Eq. (4.17a) of Ref. \cite{Arun:2008kb}. However, we present only the $x$ component here, as the applicability of the $y$ component remains to be verified. Such verification is left for future research, after obtaining the eccentric waveform of the corresponding configuration.

In the zero eccentricity limit, Eq. (\ref{eq:41}) reduces into
\begin{equation}\label{eq:42}
\mathcal{A}_{s,p}=2\mathcal{A}_c[a|\chi^x_{\rm eff}|+\sum_{i=0}^n a_i \mathcal{A}_c^i]+\mathcal{A}_c.
\end{equation}
Ref. \cite{Boyle:2011gg,Schmidt:2012rh} propose that by transforming the spin-precession gravitational waveform into a co-precessing frame and obtains a spin-aligned waveform based on a map to analytical precessing PN waveform. We believe that the operation $f_p$ for removing precession and Eq.  (\ref{eq:42}) basically represent the same meaning. Here, $f_p$ can also be understood as a similar mapping. We think Eq.(\ref{eq:42}) is an inspiration towards a more general case, that is, to establish a transformation relationship between the nonspinning or spin-aligned circular orbit waveform and spin-precessing eccentric waveform. However, it requires more numerical relativistic simulations of eccentric orbit and spin precession, which we leave for future research.

\section{results}\label{sec:III}
\subsection{2-2 mode}\label{sec:III:A}
\subsubsection{Fitting results}
In this part, we organize the fitting parameters obtained in Sec.\ref{sec:II}.
Some waveforms with large errors need to be discarded. The $q=6$ waveforms are not shown due to speculated numerical simulation data issue. As shown in FIG. \ref{FIG:5}(e), they deviate far from the behavior shown by the parameters of other waveforms. And what's more, as in panel f of the same figure, the deviation is not large, while the mass ratio $q=7$ is going larger. 

\begin{figure*}[!htbp]
\centering
\includegraphics[width=15cm,height=5cm]{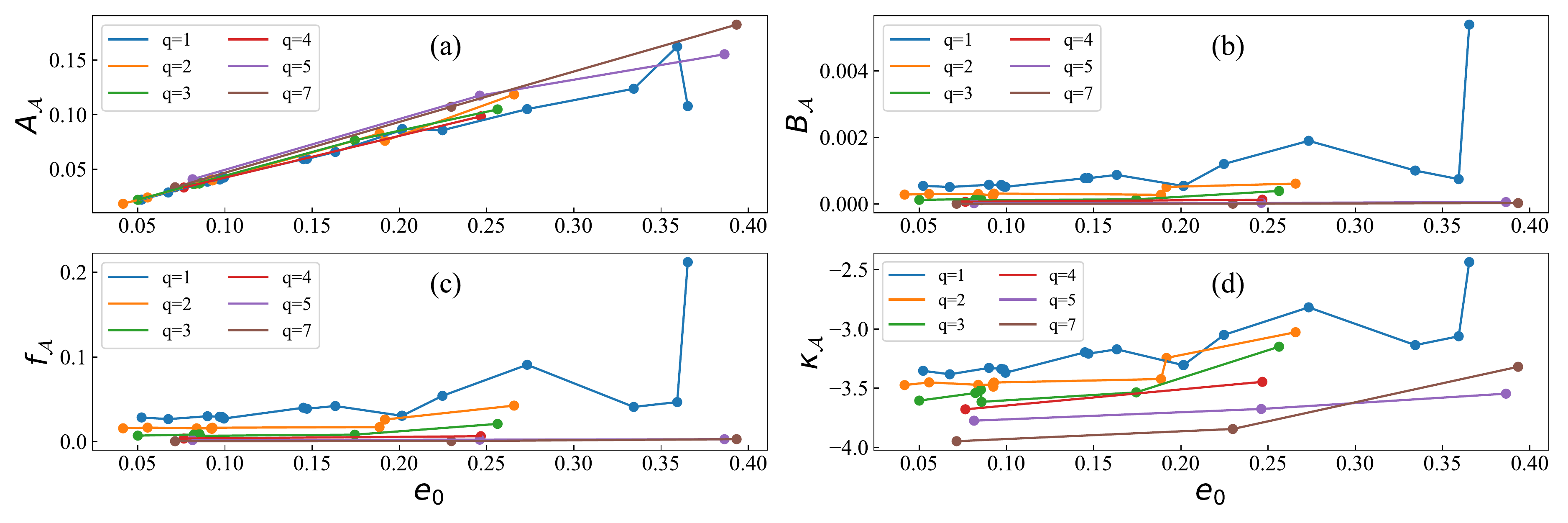}
\caption{\label{FIG:12} Fitting results of parameters $A_{\mathcal{A}}$, $B_{\mathcal{A}}$, $f_{\mathcal{A}}$, $\kappa_{\mathcal{A}}$ for different mass ratio $q$ eccentricity $e_0$ for amplitude of 2-2 mode. }
\end{figure*}

We only show the behavior here for amplitude and time period $t\in[-2000,-300]$, but the fitting parameters for frequency or other time period share the same behavior. Here, we only show fitting parameter results of the 2-2 mode to amplitude eccentricity estimator $\mathcal{A}$, and it is similar for other cases. The fitting results of parameters $A_{\mathcal{A}}$, $B_{\mathcal{A}}$, $f_{\mathcal{A}}$, $\kappa_{\mathcal{A}}$ for different mass ratio $q$ are shown in FIG. \ref{FIG:12}, where the subscript $\mathcal{A}$ represents that it is the fitting of amplitude. We take the values of $A_{\mathcal{A}}$ and $f_{\mathcal{A}}$ as positive, and $B_{\mathcal{A}}$ and $\kappa_{\mathcal{A}}$ as positive and negative. In fact, the values of $A_{\mathcal{A}}$ and $f_{\mathcal{A}}$ can be positive or negative, depending on the parity of the \textit{sine} function, while $B_{\mathcal{A}}$ and $\kappa_{\mathcal{A}}$ must be positive and negative. The sign of $A_{\mathcal{A}}$ and $f_{\mathcal{A}}$ does not affect the relationship between them and initial eccentricity $e_0$. The parameter $\varphi_{\mathcal{A}}$ has no effect on the morphological properties of the amplitude and frequency of the waveform, but only translates the frequency and amplitude on the $\mathcal{A}_c$ coordinate. That is, it does not reflect nature of initial eccentricity $e_0$ and mass ratio $q$ of the waveform. So $\varphi_{\mathcal{A}}$ is a free parameter. We also discover it has a certain periodicity.

As we see in FIG. \ref{FIG:12}, the parameters $A_{\mathcal{A}}$, $B_{\mathcal{A}}$, $f_{\mathcal{A}}$ and $\kappa_{\mathcal{A}}$ are related to the mass ratio $q$ and can be judged by the hierarchical phenomenon of the curves. In FIG. \ref{FIG:12}, we can see

(i) There is a strict proportional relationship between $A_{\mathcal{A}}$ and eccentricity $e_0$. The larger the eccentricity $e_0$ is, the larger $A_{\mathcal{A}}$ is, which comes from the relationship between the amplitude eccentricity estimator $e_{\mathcal{A}}$ and the eccentricity $e_0$. In FIG. \ref{FIG:12} (a), we see that $A_{\mathcal{A}}$ is not only related to eccentricity, but also to mass ratio $q$.

(ii) The correspondences between $B_{\mathcal{A}}$ and $f_{\mathcal{A}}$ and eccentricity $e_0$ are very similar, but their magnitudes are not the same. They are distinctly stratified at different mass ratios (see in FIG. \ref{FIG:12} (b) and (c)). And there is a slightly monotonic relationship between them and eccentricity $e_0$. However, we cannot ignore this monotonic relationship, because the accuracy of the waveform is extremely demanding on these two parameters.

(iii) Parameter $\kappa_{\mathcal{A}}$ is also related to both mass ratio $q$ and eccentricity $e_0$. When we obtain the fitting result of higher-order modes, we will find that the function of $\kappa_{\mathcal{A}}$, $f_{\mathcal{A}}$ and $B_{\mathcal{A}}$ is to adjust the Eq. (\ref{eq:28}), making it suitable for different mass ratios and higher-order modes.

(iv) The larger the eccentricity, the more spread out the points are, which means larger errors due to interpolation to measure eccentricity and fitting, as analyzed in the Sec. \ref{sec:II:C}.

These parameters have a strong dependence on eccentricity $e_0$ and mass ratio $q$. When the amount of data is relatively small, it is difficult for us to judge it as in Ref. \cite{Setyawati:2021gom}. When the amount of data is large, we can see it clearly. Since the parameters have a complex relationship not only with the mass ratio but also with the eccentricity, which makes it very difficult to cover the parameter space of $q$ and $e_0$ at the same time. Ref.\cite{Setyawati:2021gom} tries to establish the relationship between $q$, $e_0$ and parameters respectively to cover the parameter space of $q$, $e_0$, but FIG. \ref{FIG:12} shows that it is not accurate enough. We can only fix the mass ratio $q$ and then cover the eccentricity $e_0$ space. As shown in the FIG. \ref{FIG:12}, due to there are too many data points and very scattered, it is difficult to interpolate them. So we obtain a relationship between parameters $A_{\mathcal{A}}$, $B_{\mathcal{A}}$, $f_{\mathcal{A}}$, $\kappa_{\mathcal{A}}$ and $e_0$ by polynomial fitting. We use linear fit for $A_{\mathcal{A}}$. For $B_{\mathcal{A}}$, $f_{\mathcal{A}}$ and $\kappa_{\mathcal{A}}$, the results are obtained by a second order polynomial fitting, and higher orders are also possible. It is worth emphasizing that polynomial fitting does not mean that there must be a continuous monotonic relationship between these four parameters and eccentricity, which is only an approximation. If we want to get a more accurate relationship between them, we need more eccentric waveform data.

\subsubsection{Mismatch}
\begin{figure*}[!htbp]
\centering
\includegraphics[width=15cm,height=10cm]{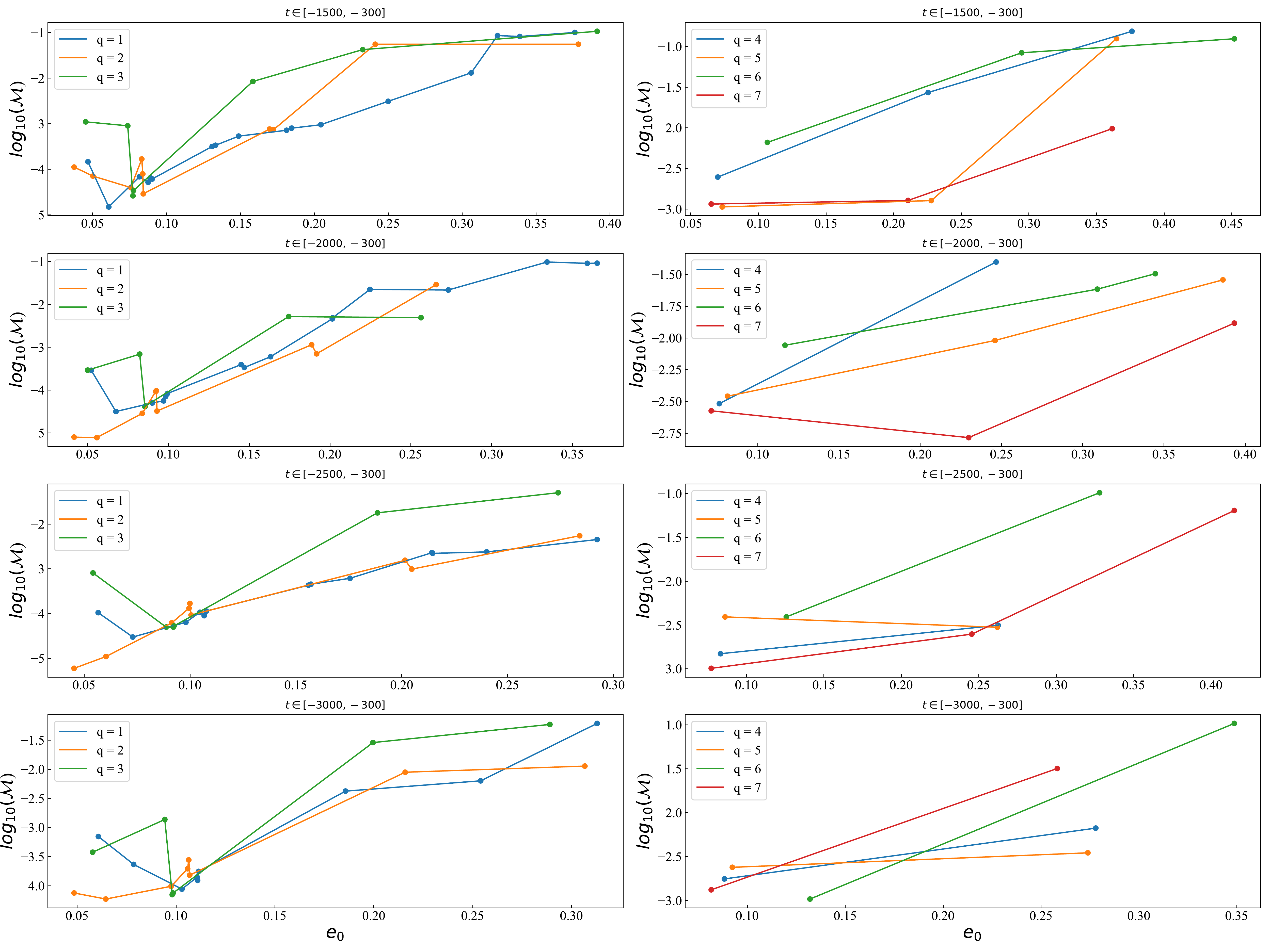}
\caption{\label{FIG:13}$\mathcal{M}$ between the waveform obtained by leave-one-out method and the test waveform in different eccentricities $e_0\in[0,0.4]$, different mass ratios $q\in[1,3]$ and different time ranges. For mass ratio $q\in[4,7]$, since there are few waveforms, we can only use the fit of the test waveform to calculate $\mathcal{M}$. }
\end{figure*}

Since our purpose is to reproduce the full numerically relativistic waveforms, they are naturally our comparison target. We use leave-one-out method, setting one waveform of all data as test data and the other waveforms as training data in order to use enough training data. First we generate fitting parameters $A_{\mathcal{A}}$, $B_{\mathcal{A}}$, $f_{\mathcal{A}}$ and $\kappa_{\mathcal{A}}$ by training data like FIG. \ref{FIG:12}. Then, correspondence between eccentricity and these fitting parameters is obtained by polynomial fitting in the FIG. \ref{FIG:12}. Next we obtain a new set of parameters $A_{\mathcal{A}}$, $B_{\mathcal{A}}$, $f_{\mathcal{A}}$ and $\kappa_{\mathcal{A}}$ through value of eccentricity $e_0$ of test waveform through this correspondence in FIG. \ref{FIG:12}. 
After we get the new fitting parameters, we can obtain the corresponding amplitude $\mathcal{A}$ and frequency $\omega$ through Eq. (\ref{eq:29}). The leave-one-out method traverses all waveform data, that is to say, each waveform has been used as test data. The new fitting parameters $A_{\omega}$, $B_{\omega}$, $f_{\omega}$ and $\kappa_{\omega}$ of frequency $\omega$ of all gravitational waveform are also generated by the same method. After this, we have to integrate the frequency to get phase by
\begin{equation}\label{eq:43}
\Phi = \int_{t_1}^{t_2}\omega dt,
\end{equation}
where ${t_1}$ and ${t_2}$ are the integral lower and upper limits. Then we can reconstruct the test waveform by Eq. (\ref{eq:19}). In order to evaluate the similarity between the test waveform and the newly reconstructed waveform, we need to calculate overlap as in Ref. \cite{Habib:2019cui}:
\begin{equation}\label{eq:44}
\mathcal{O} =\mathop{\max}\limits_{{t_0, \Phi_0,\varphi_{\mathcal{A}},\varphi_{\omega} }} \frac{\left\langle h_1, h_2\right\rangle}{\sqrt{\left\langle h_1, h_1\right\rangle\left\langle h_2, h_2\right\rangle}},
\end{equation}
where $\left\langle h_1, h_2\right\rangle$ is the inner product of waveform $h_1$ and $h_2$ defined as
\begin{equation}\label{eq:45}
\left\langle h_1, h_2\right\rangle=\left|\int_{t_{\min }}^{t_{\max }} h_1(t) h_2^*(t) d t\right|,
\end{equation}
where $h_2^*(t)$ is complex conjugate of $h_2(t)$. $t_0$ and $\Phi_0$ are given time and phase $\varphi_{\mathcal{A}}$ and $\varphi_{\omega}$ are free parameters inherited from the construction of waveform. We calculate the overlap in time domain because the waveform we constructed is a time domain waveform and each fitting parameter related to eccentricity is equivalent to related to time. At the same time, we choose a uniform Power Spectral Density (PSD) set to unity instead of the noise PSD of LIGO or other gravitational wave detectors in calculation, in order to to reflect the fitting effect in the entire time domain because we do not care about the application on detection of gravitational waves, but only care about the waveform itself. For convenience, we can also define mismatch or unfaithfulness as
\begin{equation}\label{eq:46}
 \mathcal{M} =1-\mathcal{O}.
\end{equation}
We do not calculate $\mathcal{M}$ for all time periods, but choose four typical time periods $t\in[-3000,-300]$, $t\in[-2500,-300]$, $t\in[-2000,-300]$ and $t\in[-1500,-300]$ as examples. However, we think it is common for any other continuous time period like $t\in[-1501,-300]$. Here, we need not to consider the total mass $M$, because all units have been cancelled in Eq. (\ref{eq:44}).  In FIG. \ref{FIG:13}, we show the $\mathcal{M}$ between the waveform obtained by leave-one-out method and the test waveform in different eccentricities $e_0\in[0,0.4]$, different mass ratios $q\in[1,3]$ and different time ranges presented in the left panel. For mass ratio $q\in[4,7]$, since there are few (only two or three) waveforms  which cannot generate enough fitting parameters for polynomial fitting, we can only use the fit of the test waveform to calculate $\mathcal{M}$, which is also meaningful to reflect the fitting effect of the model. We remind the reader not to confuse the two fitting procedures, one to amplitude and frequency, and the other to polynomial fitting to the fitting parameters obtained from the former.

\subsubsection{ Analysis}
$\mathcal{M}$ reflects the similarity of the waveforms. From the FIG. \ref{FIG:13}, we can see that

(i) For waveforms with different time periods, we get almost similar mismatch but there are also subtle differences. When time period is very long, such as $t\in[-3000,-300]$ or longer, or time period is short, such as $t\in[-1500,-300]$ or shorter, the fitting effect we obtained are not so good as time range $t\in[-2000,-300]$ and $t\in[-2500,-300]$. The reason for it is that when the waveform is very long, the model cannot fully capture all the overall properties of it, and when the waveform is short, the model gives a large error due to too little information given by the waveform.

(ii) For the waveforms with mass ratio $q\in[1,3]$ in the left panel in FIG. \ref{FIG:13}, we obtain a relatively small $\mathcal{M}$, but when the mass ratio is in $q\in[4,7]$ in the right panel in FIG. \ref{FIG:13}, we cannot get a small $\mathcal{M}$ which is an order of magnitude higher for all initial eccentricities even using fit of the waveform itself. It is not because the model is not suitable for mass ratio $q\in[4,7]$, but the some errors in the waveforms themselves in the RIT catalog, which is particularly obvious in the amplitude eccentricity estimator $e_{\mathcal{A}}$. So we do not show it in FIG. \ref{FIG:5} but present frequency eccentricity estimator $e_\omega$. Some $e_{\mathcal{A}}$s come from RIT catalog has a strong local ups and downs which cannot be removed using the method of filtering high-frequency noise, which is obviously caused by errors of the waveforms themselves. It can be found that the eccentric waveforms with mass ratio of 4-7 are from the RIT which used a different numerical method with SXS and have not been calibrated with other numerical relativity groups.

(iii) We can find that as the initial eccentricity $e_0$ increases, $\mathcal{M}$ becomes larger. If we only pay attention to the part with a mass ratio of 1-3 in the left panels of FIG. \ref{FIG:13}, we find that when $e_0$ is small, $\mathcal{M}$ is warped. This is because when using the leave-one-out method, there are errors in the process of generating parameters using polynomials for the waveform near the edge of the parameter. Therefore, the initial eccentricity is in the middle part to obtain a lower $\mathcal{M}$. Ignoring the initial warping part, for waveforms with mass ratio of 1-3,, when the eccentricity $e_0\in[0,0.1]$, we obtain an $\mathcal{M}$ approximately less than $10^{-4}$, and when the eccentricity $e_0\in[0.1,0.2]$, we obtain an $\mathcal{M}$ approximately less than $10^{-3}$, and when the eccentricity $e_0\in[0.2,0.3]$, we obtain an $\mathcal{M}$ approximately less than $10^{-2}$, and when the eccentricity $e_0\in[0.3,0.4]$, we obtain an $\mathcal{M}$ approximately less than $10^{-1}$. This implies that as the eccentricity becomes larger, the model fits the waveform worse, which is consistent with the conclusion we have drawn in the Sec.\ref{sec:II:D}. For the part with a mass ratio of 4-7 in the right panels of FIG. \ref{FIG:13}, the results obtained for $\mathcal{M}$ are an order of magnitude worse than 1-3.

\subsubsection{Morphology of eccentric waveform}
In this subsection, we try to explain the behavior of $\mathcal{M}$ as the initial eccentricity $e_0$ varies. There is a clear difference in morphology between the eccentric waveform and the circular orbit waveform because of modulation effects on gravitational wave amplitude and frequency oscillations due to eccentricity. The separation of BBHs is relatively close and far away at the periastron and apastron. So the amplitude and frequency of the waveform is relatively large at the periastron and relatively small at the apastron, which is the same for dominant mode in FIG. \ref{FIG:2} or higher order modes in FIG. \ref{FIG:7}. However, not only the eccentric waveform and the circular orbit waveform are morphologically different, but also the low eccentricity waveform and the high eccentricity waveform are very different in morphology.

\begin{figure}[!htbp]
\centering
\includegraphics[width=8cm,height=8cm]{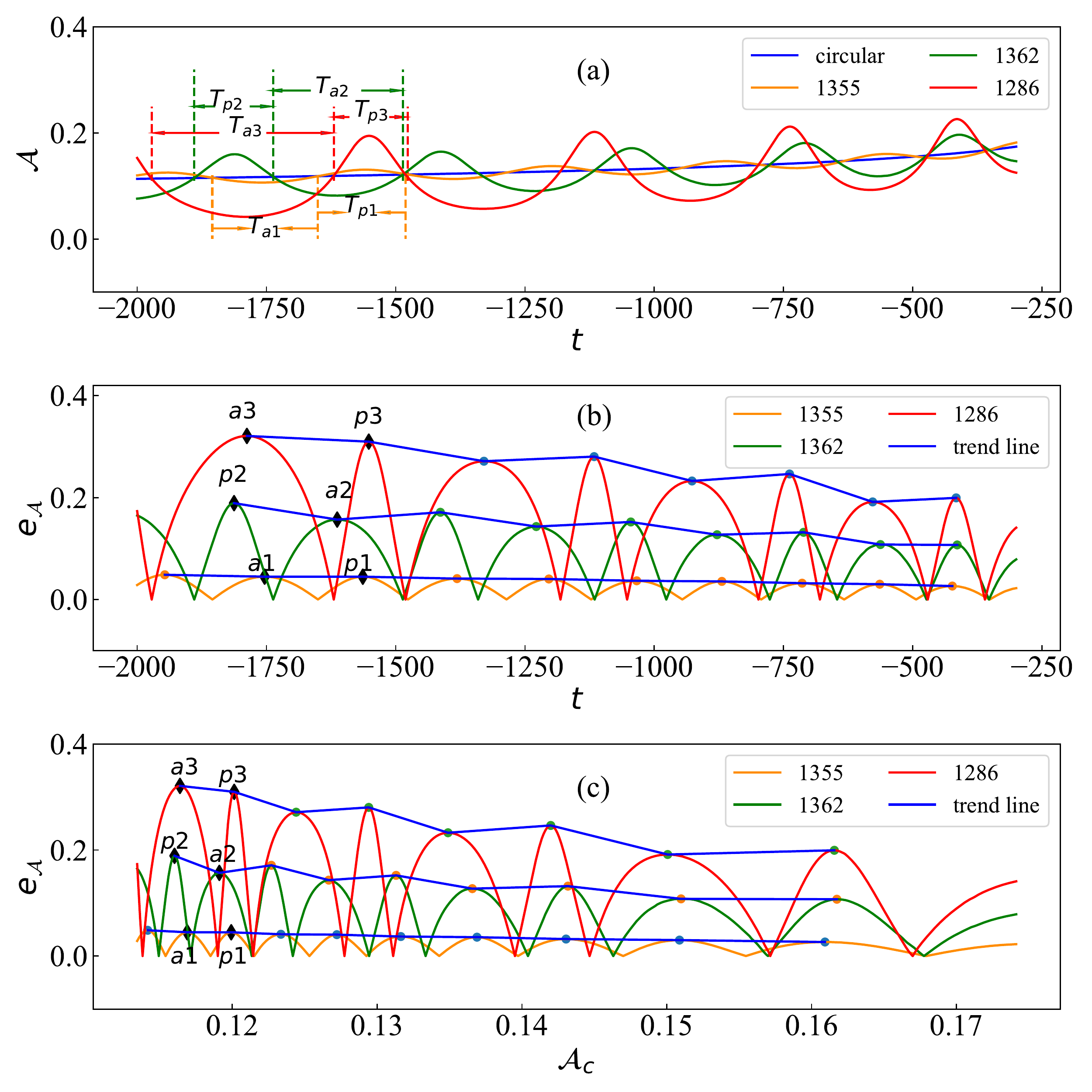}
\caption{\label{FIG:14}Amplitudes of the waveforms with mass ratio $q=1$, time period $t\in[-2000,-300]$ and initial eccentricities of $e_0=0$ (circular), $e_0=0.0522$ (1355), $e_0=0.2014$ (1362) and $e_0=0.3653$ (1286), respectively. $T_{p1}$ and $T_{a1}$ represent the time of the periastron passage and apastron passage of the first and second half cycle of the waveform 1355 based on the circular orbit waveform. $T_{p2}$, $T_{a2}$ and $T_{p3}$, $T_{a3}$ are for 1362 and 1286. From the figure, we can find that the eccentric behavior of the \textit{cosine} function is passed from panel (a) to panel (b) and then passed to panel (c).}
\end{figure}

It is difficult for us to see this morphological difference only through the comparison between the eccentric waveforms, but the circular orbit waveform and the eccentricity estimator provide us with a new perspective. The eccentricity estimator is defined as a \textit{cosine} function by Eq. (\ref{eq:9}). That is, when the eccentricity estimator deviates from the behavior of the \textit{cosine} function, measuring eccentricity by the eccentricity estimator will introduce errors \cite{Mroue:2010re}. In FIG. \ref{FIG:14}, we show the amplitudes of the waveforms with mass ratio $q=1$, time period $t\in[-2000,-300]$ and initial eccentricities of $e_0=0$ (circular), $e_0=0.0522$ (1355), $e_0=0.2014$ (1362) and $e_0=0.3653$ (1286), respectively. $T_{p1}$ and $T_{a1}$ represent the time of the periastron passage and apastron passage of the first and second half cycle of the waveform 1355 based on the circular orbit waveform. $T_{p2}$, $T_{a2}$ and $T_{p3}$, $T_{a3}$ are for 1362 and 1286. From FIG. \ref{FIG:14} (a), we can find that, the greater the eccentricity of the waveform, the stronger the oscillation in amplitude. As the eccentricity increases, the periastron passage and apastron passage of the waveform gradually show different behaviors. The former becomes sharper and the latter becomes smoother. We can get the ratio $T_{a3}/T_{p3}>T_{a2}/T_{p2}>T_{a2}/T_{p2}$, which means that the time of the apastron passage is getting longer and longer than the time of the periastron passage. 

It is not obvious to describe the deviation of the eccentricity estimator from the cosine function, because the eccentricity is a function of time. When using the eccentricity estimator to estimate the eccentricity, we obtain its eccentricity by taking the amplitude of the eccentricity estimator $e_{\mathcal{A}}$. This approach relies on the fact that the pericentric and apocentric values of the eccentricity estimator must decay in the same behavior, rather than there is an obvious hierarchical behavior. This hierarchical behavior shows the eccentricity estimator deviating from the sinusoidal decay. Behavior in amplitude in FIG. \ref{FIG:14}  (a) is passed to the associated amplitude eccentricity estimator. In order to show this effect, in FIG. \ref{FIG:14} (b), we take the absolute value of the eccentricity estimator, so that periastron and apastron are at the same level, and then we connect all the points to draw a trend line, where diamonds $a1$, $p1$, etc. are the corresponding periastron and apastron in the FIG. \ref{FIG:14} (a). If the eccentricity estimator had no deviating cosine behavior, then all periastron and apastron values $a_1$, $a_2$... $a_n$, and $p_1$, $p_2$... $p_n$ (we only marked two of them as examples)  would decrease in a line that approximately coincides, since eccentricity is a decreasing function of time. Conversely, if it deviates from the cosine function, the line connecting the two values will not coincide and will behave like a polyline. The ups and downs of these trend lines indicate how far the eccentricity estimator deviates from the \textit{cosine} function. The trend line of waveform 1355 is roughly a straight line, implying that it does not deviate from \textit{cosine} behavior, but the trend line of waveform 1286 is an obvious broken line, implying that it deviates from \textit{cosine} behavior. 

Due to the monotonic relationship between the amplitude of the circular orbit and time, the behavior of the amplitude eccentricity estimator with respect to time will be passed to the behavior of the  amplitude eccentricity estimator with respect to the amplitude of the circular orbit(see in FIG. \ref{FIG:14} (c)). We can derive the behavior of the Eq. (\ref{eq:28}) by magnitude analysis. From the FIG. \ref{FIG:12}, we get $B\sim10^{-3}$, ${X_c}^{\kappa}\sim10^{3}$ and $e^{B X_c^\kappa}\sim1$. So the overall behavior of Eq. (\ref{eq:29}) is a \textit{sine} or \textit{cosine} function which is similar to FIG. \ref{FIG:4} (a). The trend lines in FIG. \ref{FIG:14} (c) show that 1355 does not deviate from sinusoidal behavior, while 1286 does. All in all, the degree of deviation of the eccentricity estimator from the sinusoidal function determines the scope of application of the phenomenological fitting model. The greater the eccentricity, the greater the deviation of the eccentricity estimator from the sinusoidal behavior, which also determines the model cannot be used for high eccentricity. This result is consistent with our previous analysis of overlapping. As we can see in the FIG. \ref{FIG:13}, different eccentricities give different fitting effects, so when we want to obtain higher accuracy, we generally need to keep the eccentricity at different intervals in $[0,0.4]$.

\subsection{Other situations}\label{sec:III:B}
\subsubsection{Higher-order modes}
Compared with the 2-2 mode, the high-order modes have a large difference in amplitude and frequency, which leads to different magnitudes of parameters, but their related behavior with eccentricity and mass ratio is the same as the 2-2 mode. Similarly, we can also obtain it according to the method in the Sec. \ref{sec:III:A}. Here we will not go into details. 

\begin{figure}[!htbp]
\centering
\includegraphics[width=8cm,height=4cm]{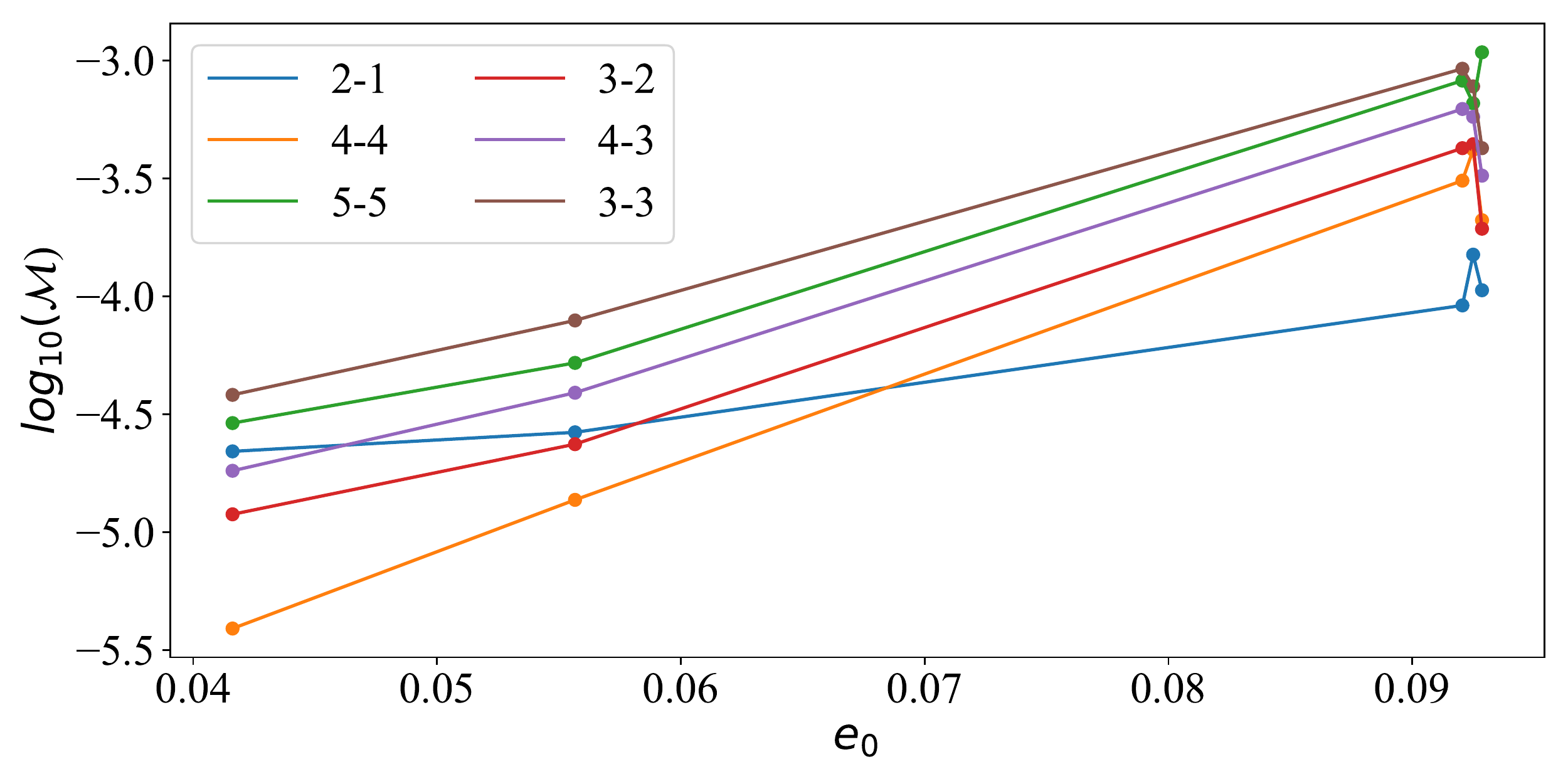}
\caption{\label{FIG:15} Mismatch $\mathcal{M}$ for 3-3, 2-1, 4-4, 5-5, 3-2, 4-3 modes with mass ratio $q=2$, eccentricity $e_0\in[0,0.1]$ and  time period $t\in[-2000,-300]$.}
\end{figure}

The calculation of mismatch $\mathcal{M}$ can be obtained through the Eq. (\ref{eq:46}). We take fitting of the high-order modes 3-3, 2-1, 4-4, 5-5, 3-2, 4-3 mode with mass ratio $q=2$, eccentricity $e_0\in[0,0.1]$ and  time period $t\in[-2000,-300]$ as an example to show the mismatch $\mathcal{M}$ we get (see in FIG. \ref{FIG:15}). The situation is similar for other parameters. Here we only use fitting to calculate $\mathcal{M}$ because there is too little data for the high-order modes of the eccentric waveforms. When we get enough high-order modes data of the eccentric waveforms, we can also do the same procedure as the Sec. \ref{sec:III:A}. From FIG. \ref{FIG:15}, we can see that the mismatch $\mathcal{M}$ of high-order modes and the 2-2 mode share roughly the same behavior with the eccentricity. Therefore, this model is able to fit the higher order modes very well.

\subsubsection{Spin-aligned}
\begin{figure}[!htbp]
\centering
\includegraphics[width=8cm,height=4cm]{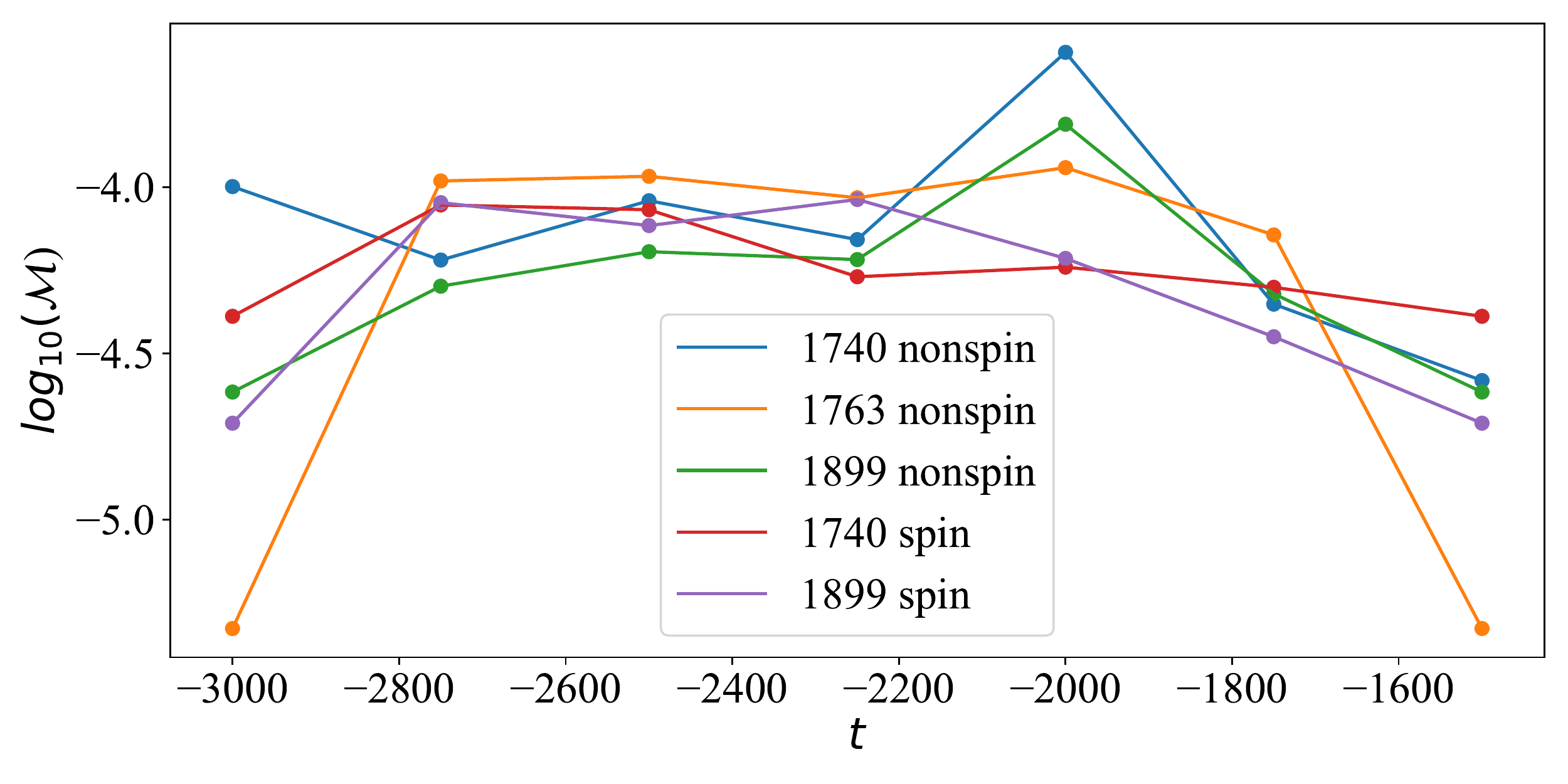} 
\caption{\label{FIG:16}Mismatch $\mathcal{M}$ of different time range with mass ratio $q=1$. RIT:eBBH:1740, RIT:eBBH:1763 and RIT:eBBH:1899 are come from circular orbit waveforms without spin, and the other two RIT:eBBH:1763 and RIT:eBBH:1899 are come from circular orbit waveforms with spin. }
\end{figure}

As stated in the Sec. \ref{sec:II:F}, the waveform of spin aligned can be obtained in two ways, one is the circular orbit waveform with spin, and the other is the circular orbit waveform without spin. We can fit waveforms by Eq. (\ref{eq:29}) or Eq. (\ref{eq:32}). Since the number of eccentric spinning waveforms is few, and it is not easy to find an eccentric waveform whose magnitude and direction of component spins are exactly the same as a circular orbital waveform, we can only present the fitting effect for two cases here. We show mismatch $\mathcal{M}$ of different time range with mass ratio $q=1$ in the FIG. \ref{FIG:16} in which RIT:eBBH:1740, RIT:eBBH:1763 and RIT:eBBH:1899 are for circular orbit waveforms without spin, and the other two RIT:eBBH:1763 and RIT:eBBH:1899 are for circular orbit waveforms with spin. 
Due to the limited number of spin-aligned waveforms, we cannot demonstrate the variation of $\mathcal{M}$ with eccentricity $e_0$. Instead, we take the time range from [-3000, -300] to [-1500, -300], where we choose the middle time range every 250, such as [-2750, -300], [-2500, -300], and so on.
Since the initial eccentricities of the waveforms in FIG. \ref{FIG:16} are similar and located within $e_0\in[0,0.1]$, we observe that their corresponding $\mathcal{M}$ values are approximately less than $10^{-4}$. This observation suggests that the proposed model can fit these waveforms well.

\subsubsection{Spin-precession}
\begin{figure}[!htbp]
\centering
\includegraphics[width=8cm,height=8cm]{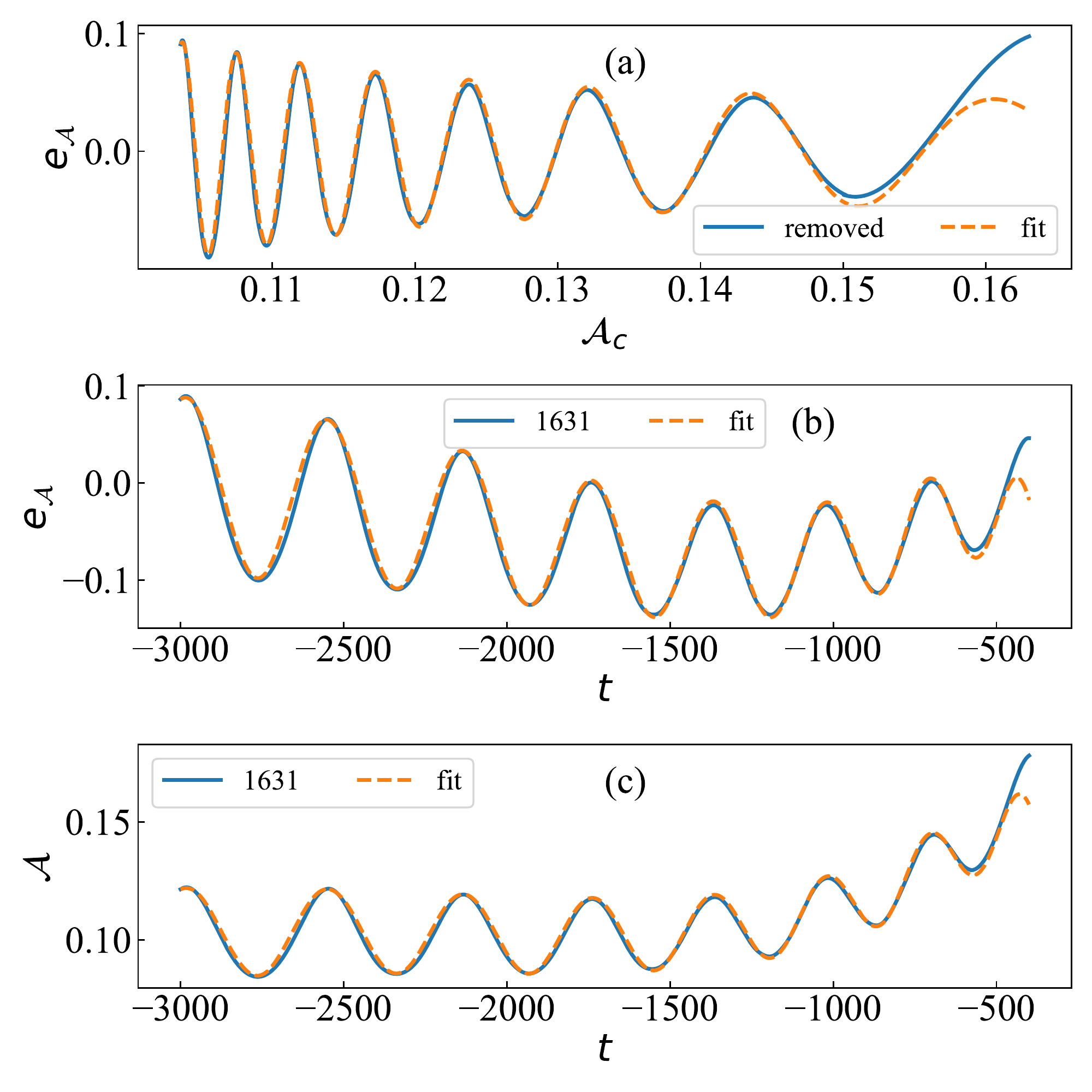} 
\caption{\label{FIG:17}In panel (a), the blue solid line represents the part of the waveform RIT:eBBH:1361 in the time range $t\in[-3000,-300]$, which is named ``removed", because we removed the effective spin and precession effects inside it. The dark orange dashed line represents a new amplitude eccentricity estimator we ``fit" . In panel (b), the blue solid line represents the ``fit" amplitude eccentricity estimator of waveform that we add the effective spin effect $-g=0.3$ and the precession effect $f_p$. The dark orange dashed line represents the 1631 amplitude eccentricity estimator. In panel (c), The dark orange dashed line is the ``fit" the amplitude of a waveform with eccentricity and spin precession by the Eq. (\ref{eq:41}). The blue solid line represents the amplitude of waveform 1631.}
\end{figure}

\begin{figure*}[!htbp]
\centering
\includegraphics[width=15cm,height=2.5cm]{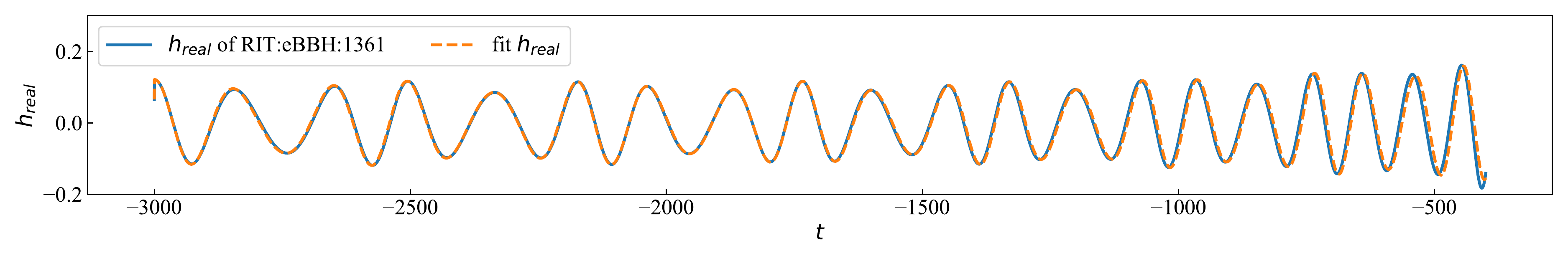}
\caption{\label{FIG:18} A phenomenological comparison between  $h_{real}$ that we ``fit" and original eccentric spin-precessing waveform  RIT:eBBH:1361 for 2-2 mode.}
\end{figure*}

The eccentric spin-precessing BBHs merger is one of the most complicated cases in numerical relativity, and there are few studies on it. Their waveforms are interspersed with multiple complex effects and are difficult to analyze. In the Sec. \ref{sec:II:G}, we propose a method for separating the eccentric spin-precessing waveform into different effects as in Eq. (\ref{eq:39}). Now, we select 2-2 mode of the waveform RIT:eBBH:1361 in the time period $t\in[-3000,-300]$. We can test this model by doing a fit to the nonspinning waveform in FIG. \ref{FIG:11} (b). Then we obtain the values of each parameter $A_\mathcal{A}=0.03459$, $B_\mathcal{A}=0.000786$, $f_\mathcal{A}=0.04617$, $\kappa_\mathcal{A}=-3.15750$. Next we  obtain an eccentricity estimator with mass ratio $q=1$, no spin and precession via the Eq. (\ref{eq:39}). The results are shown in FIG. \ref{FIG:17} (a), where the blue solid line is named ``removed", because we removed the effective spin and precession effects inside it, and the dark orange dashed line is what we ``fit". Then we add the effective spin effect $-g=0.3$ and the precession effect $f_p$ to the ``fit" waveform to obtain an eccentricity estimator with spin-precession through the Eq. (\ref{eq:40}), which precisely recovers the characteristics of the eccentricity estimator of original waveform 1631(see in FIG. \ref{FIG:17}(b)). Finally, we get the amplitude of a waveform with eccentricity and spin precession by the Eq. (\ref{eq:41}), which has a good overlap with the amplitude of waveform 1631(see in FIG. \ref{FIG:17} (c)). For frequency $\omega$, the manipulations are the same. Substitute the resulting frequency $\omega$ and amplitude $\mathcal{A}$ into the Eq. (\ref{eq:19}) and compare with the original waveform RIT:eBBH:1361 we get a mismatch $\mathcal{M}=0.0016$, which means that we accurately and self-consistently reproduce the original waveform 1631. A phenomenological comparison between  $h_{real}$ we ``fit" and  original waveform  RIT:eBBH:1361 is shown in FIG.\ref{FIG:18}, where $h_{real}$ is the real part of $h$.

\begin{figure}[!htbp]
\centering
\includegraphics[width=8cm,height=4cm]{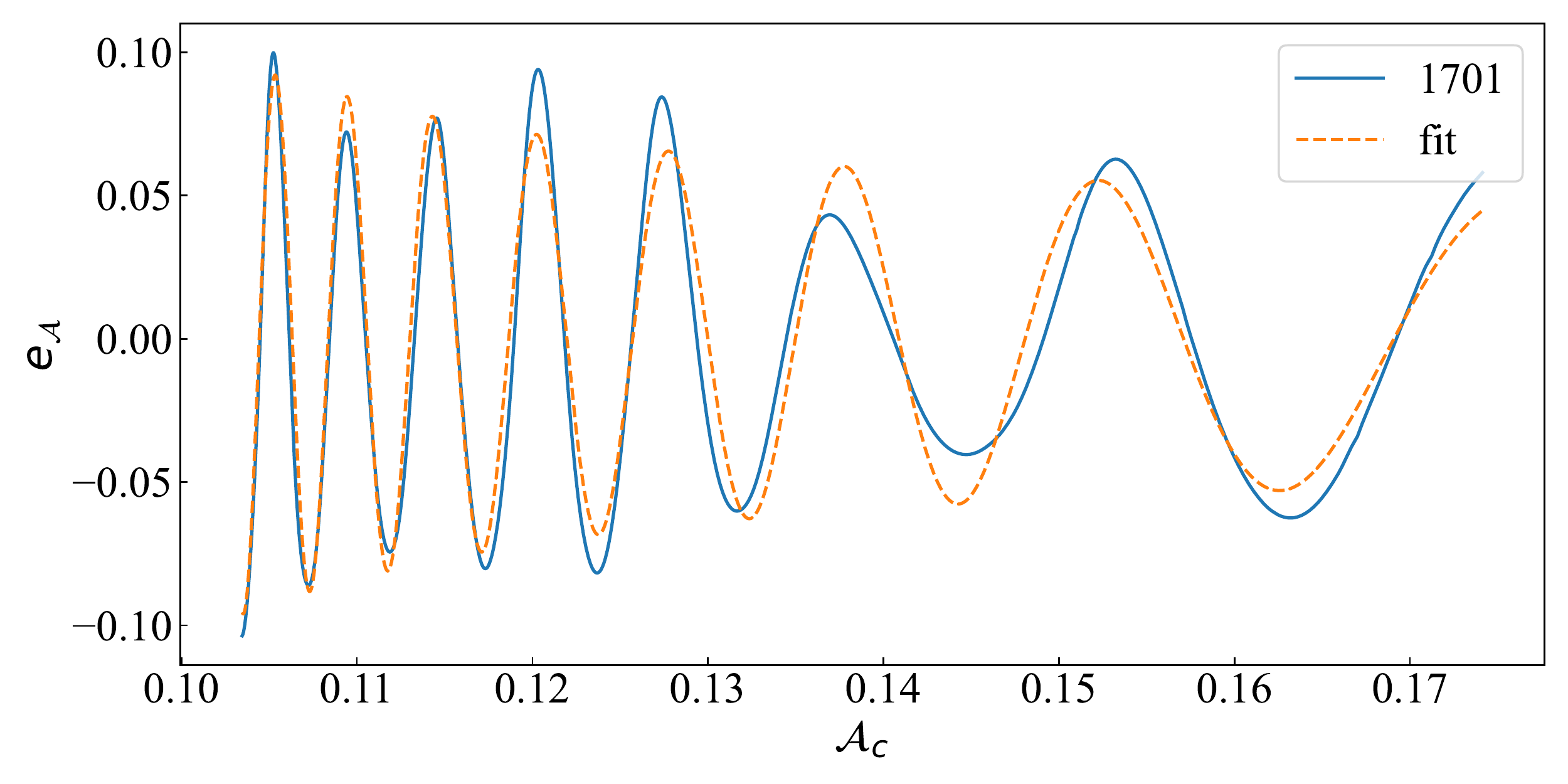} 
\caption{\label{FIG:19}The blue solid line represents amplitude eccentricity estimator of waveform 1701, and the dark orange dashed line represents a forced fit for it.}
\end{figure}

Not all eccentric spin strong precession waveforms have a simple precession effect like RIT:eBBH:1361. Waveform RIT:eBBH:1701 has an effective spin of zero due to the spin opposite sign of BBHs, leading to $g=0$. If we assume that the precession effect and the eccentricity effect are independent of each other, then the parameter $f$ only depends on the eccentricity. So we can force fit its eccentricity estimator to try to obtain its precession effect(see in FIG. \ref{FIG:19}). As in the previous case, subtracting the fit, we get the precession effect which is much more complicated than the precession effect in RIT:eBBH:1361.


\section{Conclusion and Outlook}\label{sec:IV}
The eccentricity and spin of gravitational waves can reflect the dynamics of BBHs merger. However, there are very few public numerical relativity simulations with eccentricity. \citet{Setyawati:2021gom} propose a novel method to convert quasi-circular orbit waveforms into eccentric ones, but their method is very limited because of the small range of parameters. We explore the origin of this phenomenological model and its possible extensions. Based on the eccentric waveforms of the SXS catalog, we add some eccentric waveforms of the RIT catalog, including the waveforms of mass ratio $q\in[1,7]$ , eccentricity $e_0\in[0,0.4]$ , spin alignment, and spin precession, greatly expanding the parameter space. We find that after setting the fixed constant parameter to the variable parameter $\kappa$ as in Eq. (\ref{eq:29}), the applicability of the model becomes wider, and can be applied to the case of mass ratio $q\in[1,7]$, eccentricity up to $e_0=0.4$, time period up to $t\in[-12000,-300]$, high-order modes and spin alignment. We use the leave-one-out method to verify this model. For $q$ 1-3, $e_0\in[0,0.1]$, it gives a mismatch approximately less than $10^{-4}$, for $e_0\in[0.1,0.2]$, less than $10^{-3}$, for $e_0\in[0.2,0.3]$, less than $10^{-2}$, for $e_0\in[0.3,0.4]$, less than $10^{-1}$. And an order of magnitude worse for mass ratio 4-7. The reason for these phenomenons is that the larger the eccentricity, the larger the deviation of the eccentricity estimator from the \textit{cosine} function due to the large change in the morphology of the eccentric waveform, and the worse the fitting effect of the model. We also try to introduce an approximate spin-aligned effect. After adding a shift parameter as in Eq. (\ref{eq:32}), the model can convert a nonspinning circular waveform into a spin-aligned eccentric waveform. That is, the spin-aligned effect can be approximately added to nonspinning waveform. For some waveforms with relatively simple and obvious precession effects strong precession, we can separate the precession effects and self-consistently restore the original waveform using the measured initial eccentricity. Finally, the spin precession waveform with eccentricity can be regarded as the superposition of various effects including eccentricity, effective spin and precession, which is very novel and simple. We can also obtain models of complex precession phenomena by the model. We believe that this \textit{phenomenological and universal} relationship can not only help us to generate fast and accurate gravitational waveforms with eccentricity and spin precession, but also put up with a new perspective for understanding eccentricity, spin and precession effects of the waveform BBHs dynamics.

As outlined in the main text, the gravitational waveform can be obtained by following a specific process based on a set of parameters. Firstly, a selection is made of a particular set of parameters including the mass ratio $q$, eccentricity $e_0$, time period, and effective spin $\chi_{\rm eff}$. Secondly, utilizing FIG. \ref{FIG:12}, four fitting parameters $A$, $B$, $f$, and $\kappa$, corresponding to the mass ratio $q$ and initial eccentricity $e_0$, are obtained. Thirdly, these four parameters are applied to Eq. (\ref{eq:30}), resulting in the determination of the amplitude $\mathcal{A}$ and frequency $\omega$, which vary with time. In the case of spin or precession, Eq. (\ref{eq:34}) and Eq. (\ref{eq:41}) should be utilized as a replacement for Eq. (\ref{eq:30}). Fourthly, $\mathcal{A}$ and $\omega$ are substituted into Eq. (\ref{eq:19}), while the phase $\Phi$ in Eq. (\ref{eq:19}) is the integration of Eq. (\ref{eq:20}). The resulting expression is the gravitational waveform $h$. It should be noted that this process is currently only applicable to a specific integer mass ratio and strong precession effects, and is not suitable for considering general precession effects.

Due to the small amount of waveform data used in this work and the lack of generality in the cases of spin alignment and spin precession, we only study some special cases for them and using some approximation. It will be necessary to include more general spin precession effects in this phenomenological model. We hope that there will be more and more numerical relativity simulations with eccentricity and spin precession in the future, which will make this phenomenological model richer and more accurate.

Once a large coverage on the parameters including mass ratio, eccentricity, and spins can be considered in the phenomenological relationship, we will be able to construct a large amount of waveforms easily. This could be used as the waveform template in searching GWs by LIGO/Virgo \citep{Abbott_2009} or by the upcoming next generation gravitational wave detectors, such as the Einstein Telescope \citep{2010CQGra..27s4002P}. As the waveforms are scalable to any mass, it can also be applied to the merging of galactic center binary black holes, which happens in the early universe, and could be detectable by the space borne gravitational wave missions, such as Laser Interferometer Space Antenna (LISA) \citep{2007CQGra..24R.113A}, and Tianqin \citep{2016CQGra..33c5010L}.

\begin{acknowledgments}
We express our sincere gratitude to the anonymous referee for providing us with an extensive list of comments and suggestions, which have significantly improved the quality of our manuscript. The authors are very grateful to the SXS collaboration and the RIT collaboration for the numerical simulation of eccentric BBHs mergers, and thanks to Yun-Gui Gong, Xiao-Lin Liu, Ying-Yan Li, Chao Zhang, Qingwen Wu, and Shi-Yan Tian for their helpful discussions.
This work is in part supported by the National Key R\&D Program of China (2022SKA0130103, 2021YFC2203100), and by the National Natural Science Foundation of China (Grant Nos. 12041306 and U1931203). We also acknowledge the science research grants from the China Manned Space Project with No. CMS-CSST-2021-B11. The computation is partly completed in the HPC Platform of Huazhong University of Science and Technology. The languages was polished by ChatGPT during the revision of the draft.
\end{acknowledgments}

\bibliographystyle{apsrev4-2}

\bibliography{ref}

\clearpage

\appendix
\section{RIT and SXS waveforms}\label{App:A}
\begin{table*}[!t]
\caption{\label{tab:I}Eccentric and circular orbit waveforms from the RIT catalog, which contains simulation number, type, mass ratio $q\in[1,7]$, initial separation $\operatorname{sep}_{ini}$, dimensionless spin component, effective spin, reference eccentricity $e_{ref}$ and merge time $t_{merge}$.}
    \centering
    \begin{tabular}{|r|r|r|r|r|r|r|r|r|r|r|r|r|r|}
    \hline
        case & simulation & type & $sep_{ini}$ & $q$ & $\chi_{1x}$ & $\chi_{1y}$ & $\chi_{1z}$ & $\chi_{2x}$ & $\chi_{2y}$ & $\chi_{2z}$ & $\chi_{eff}$ & $e_{ref}$ & $t_{merge}/M$ \\ \hline
        1 & RIT:eBBH:1282 & Nonspinning & 24.64  & 1  & 0.0  & 0.0  & 0.0  & 0.0  & 0.0  & 0.0  & 0.000  & 0.1900  & 11764  \\ \hline
        2 & RIT:eBBH:1283 & Nonspinning & 24.64  & 1  & 0.0  & 0.0  & 0.0  & 0.0  & 0.0  & 0.0  & 0.000  & 0.2775  & 6524  \\ \hline
        3 & RIT:eBBH:1284 & Nonspinning & 24.64  & 1  & 0.0  & 0.0  & 0.0  & 0.0  & 0.0  & 0.0  & 0.000  & 0.3276  & 4430  \\ \hline
        4 & RIT:eBBH:1285 & Nonspinning & 24.64  & 1  & 0.0  & 0.0  & 0.0  & 0.0  & 0.0  & 0.0  & 0.000  & 0.3600  & 3368  \\ \hline
        5 & RIT:eBBH:1287 & Nonspinning & 24.64  & 1  & 0.0  & 0.0  & 0.0  & 0.0  & 0.0  & 0.0  & 0.000  & 0.3916  & 2508  \\ \hline
        6 & RIT:eBBH:1286 & Nonspinning & 24.64  & 1  & 0.0  & 0.0  & 0.0  & 0.0  & 0.0  & 0.0  & 0.000  & 0.3994  & 2330  \\ \hline
        7 & RIT:eBBH:1289 & Nonspinning & 24.64  & 1  & 0.0  & 0.0  & 0.0  & 0.0  & 0.0  & 0.0  & 0.000  & 0.4071  & 2141  \\ \hline
        8 & RIT:eBBH:1288 & Nonspinning & 24.64  & 1  & 0.0  & 0.0  & 0.0  & 0.0  & 0.0  & 0.0  & 0.000  & 0.4148  & 1993  \\ \hline
        9 & RIT:eBBH:1291 & Nonspinning & 24.64  & 1  & 0.0  & 0.0  & 0.0  & 0.0  & 0.0  & 0.0  & 0.000  & 0.4224  & 1822  \\ \hline
        10 & RIT:eBBH:1290 & Nonspinning & 24.64  & 1  & 0.0  & 0.0  & 0.0  & 0.0  & 0.0  & 0.0  & 0.000  & 0.4300  & 1681  \\ \hline
        11 & RIT:eBBH:1293 & Nonspinning & 24.64  & 1  & 0.0  & 0.0  & 0.0  & 0.0  & 0.0  & 0.0  & 0.000  & 0.4375  & 1558  \\ \hline
        12 & RIT:eBBH:1292 & Nonspinning & 24.64  & 1  & 0.0  & 0.0  & 0.0  & 0.0  & 0.0  & 0.0  & 0.000  & 0.4450  & 1410  \\ \hline
        13 & RIT:eBBH:1740 & Spin-aligned & 24.71  & 1  & 0.0  & 0.0  & -0.5  & 0.0  & 0.0  & -0.5  & -0.500  & 0.1900  & 10094  \\ \hline
        14 & RIT:eBBH:1763 & Spin-aligned & 24.75  & 1  & 0.0  & 0.0  & -0.8  & 0.0  & 0.0  & -0.8  & -0.800  & 0.1900  & 9054  \\ \hline
        15 & RIT:eBBH:1899 & Spin-aligned & 24.69  & 1  & 0.0  & 0.0  & 0.0  & 0.0  & 0.0  & -0.8  & -0.400  & 0.1900  & 10588  \\ \hline
        16 & RIT:eBBH:1741 & Spin-aligned & 24.71  & 1  & 0.0  & 0.0  & -0.5  & 0.0  & 0.0  & -0.5  & -0.500  & 0.3600  & 2286  \\ \hline
        17 & RIT:eBBH:1764 & Spin-aligned & 24.75  & 1  & 0.0  & 0.0  & -0.8  & 0.0  & 0.0  & -0.8  & -0.800  & 0.3600  & 1620  \\ \hline
        18 & RIT:eBBH:1900 & Spin-aligned & 24.69  & 1  & 0.0  & 0.0  & 0.0  & 0.0  & 0.0  & -0.8  & -0.400  & 0.3600  & 2483  \\ \hline
        19 & RIT:eBBH:1786 & Spin-aligned & 24.59  & 1  & 0.0  & 0.0  & 0.5  & 0.0  & 0.0  & 0.5  & 0.500  & 0.4375  & 2448  \\ \hline
        20 & RIT:eBBH:1807 & Spin-aligned & 24.56  & 1  & 0.0  & 0.0  & 0.8  & 0.0  & 0.0  & 0.8  & 0.800  & 0.4375  & 2985  \\ \hline
        21 & RIT:eBBH:1828 & Spin-aligned & 24.60  & 1  & 0.0  & 0.0  & 0.0  & 0.0  & 0.0  & 0.8  & 0.400  & 0.4375  & 2275  \\ \hline
        22 & RIT:eBBH:1787 & Spin-aligned & 24.59  & 1  & 0.0  & 0.0  & 0.5  & 0.0  & 0.0  & 0.5  & 0.500  & 0.4671  & 1866  \\ \hline
        23 & RIT:eBBH:1808 & Spin-aligned & 24.56  & 1  & 0.0  & 0.0  & 0.8  & 0.0  & 0.0  & 0.8  & 0.800  & 0.4671  & 2349  \\ \hline
        24 & RIT:eBBH:1829 & Spin-aligned & 24.60  & 1  & 0.0  & 0.0  & 0.0  & 0.0  & 0.0  & 0.8  & 0.400  & 0.4671  & 1720  \\ \hline
        25 & RIT:eBBH:1631 & Spin-precessing & 24.62  & 1  & 0.7  & 0.0  & 0.0  & 0.7  & 0.0  & 0.0  & 0.700  & 0.1900  & 12012  \\ \hline
        26 & RIT:eBBH:1701 & Spin-precessing & 24.64  & 1  & -0.7  & 0.0  & 0.0  & 0.7  & 0.0  & 0.0  & 0.000  & 0.1900  & 11759  \\ \hline
        27 & RIT:eBBH:1632 & Spin-precessing & 24.62  & 1  & 0.7  & 0.0  & 0.0  & 0.7  & 0.0  & 0.0  & 0.700  & 0.2775  & 6726  \\ \hline
        28 & RIT:eBBH:1702 & Spin-precessing & 24.64  & 1  & -0.7  & 0.0  & 0.0  & 0.7  & 0.0  & 0.0  & 0.000  & 0.2775  & 6554  \\ \hline
        29 & RIT:eBBH:1633 & Spin-precessing & 24.62  & 1  & 0.7  & 0.0  & 0.0  & 0.7  & 0.0  & 0.0  & 0.700  & 0.3600  & 3488  \\ \hline
        30 & RIT:eBBH:1703 & Spin-precessing & 24.64  & 1  & -0.7  & 0.0  & 0.0  & 0.7  & 0.0  & 0.0  & 0.000  & 0.3600  & 3382  \\ \hline
        31 & RIT:eBBH:1634 & Spin-precessing & 24.62  & 1  & 0.7  & 0.0  & 0.0  & 0.7  & 0.0  & 0.0  & 0.700  & 0.4375  & 1633  \\ \hline
        32 & RIT:eBBH:1704 & Spin-precessing & 24.64  & 1  & -0.7  & 0.0  & 0.0  & 0.7  & 0.0  & 0.0  & 0.000  & 0.4375  & 1562  \\ \hline
        33 & RIT:eBBH:1883 & Spin-aligned & 24.70  & 2  & 0.0  & 0.0  & 0.0  & 0.0  & 0.0  & -0.8  & -0.533  & 0.3600  & 2257  \\ \hline
        34 & RIT:eBBH:1422 & Nonspinning & 24.63  & 2  & 0.0  & 0.0  & 0.0  & 0.0  & 0.0  & 0.0  & 0.000  & 0.1900  & 13074  \\ \hline
        35 & RIT:eBBH:1423 & Nonspinning & 24.63  & 2  & 0.0  & 0.0  & 0.0  & 0.0  & 0.0  & 0.0  & 0.000  & 0.3600  & 3714  \\ \hline
        36 & RIT:eBBH:1424 & Nonspinning & 24.63  & 2  & 0.0  & 0.0  & 0.0  & 0.0  & 0.0  & 0.0  & 0.000  & 0.4375  & 1667  \\ \hline
        37 & RIT:eBBH:1862 & Spin-aligned & 24.70  & 3  & 0.0  & 0.0  & 0.0  & 0.0  & 0.0  & -0.8  & -0.600  & 0.3600  & 2209  \\ \hline
        38 & RIT:eBBH:1468 & Nonspinning & 24.62  & 3  & 0.0  & 0.0  & 0.0  & 0.0  & 0.0  & 0.0  & 0.000  & 0.1900  & 15114  \\ \hline
        39 & RIT:eBBH:1469 & Nonspinning & 24.62  & 3  & 0.0  & 0.0  & 0.0  & 0.0  & 0.0  & 0.0  & 0.000  & 0.3600  & 4249  \\ \hline
        40 & RIT:eBBH:1470 & Nonspinning & 24.62  & 3  & 0.0  & 0.0  & 0.0  & 0.0  & 0.0  & 0.0  & 0.000  & 0.4375  & 1854  \\ \hline
        41 & RIT:eBBH:1491 & Nonspinning & 24.61  & 4  & 0.0  & 0.0  & 0.0  & 0.0  & 0.0  & 0.0  & 0.000  & 0.1900  & 17835  \\ \hline
        42 & RIT:eBBH:1492 & Nonspinning & 24.61  & 4  & 0.0  & 0.0  & 0.0  & 0.0  & 0.0  & 0.0  & 0.000  & 0.3600  & 4878  \\ \hline
        43 & RIT:eBBH:1493 & Nonspinning & 24.61  & 4  & 0.0  & 0.0  & 0.0  & 0.0  & 0.0  & 0.0  & 0.000  & 0.4375  & 2102  \\ \hline
        44 & RIT:eBBH:1514 & Nonspinning & 24.60  & 5  & 0.0  & 0.0  & 0.0  & 0.0  & 0.0  & 0.0  & 0.000  & 0.1900  & 15781  \\ \hline
        45 & RIT:eBBH:1515 & Nonspinning & 24.60  & 5  & 0.0  & 0.0  & 0.0  & 0.0  & 0.0  & 0.0  & 0.000  & 0.3600  & 5539  \\ \hline
        46 & RIT:eBBH:1516 & Nonspinning & 24.60  & 5  & 0.0  & 0.0  & 0.0  & 0.0  & 0.0  & 0.0  & 0.000  & 0.4375  & 2496  \\ \hline
        47 & RIT:eBBH:1517 & Nonspinning & 24.60  & 5  & 0.0  & 0.0  & 0.0  & 0.0  & 0.0  & 0.0  & 0.000  & 0.4671  & 1601  \\ \hline
        48 & RIT:eBBH:1537 & Nonspinning & 24.59  & 6  & 0.0  & 0.0  & 0.0  & 0.0  & 0.0  & 0.0  & 0.000  & 0.1900  & 6268  \\ \hline
        49 & RIT:eBBH:1538 & Nonspinning & 24.59  & 6  & 0.0  & 0.0  & 0.0  & 0.0  & 0.0  & 0.0  & 0.000  & 0.3600  & 3200  \\ \hline
        50 & RIT:eBBH:1539 & Nonspinning & 24.59  & 6  & 0.0  & 0.0  & 0.0  & 0.0  & 0.0  & 0.0  & 0.000  & 0.4375  & 2213  \\ \hline
        51 & RIT:eBBH:1540 & Nonspinning & 24.59  & 6  & 0.0  & 0.0  & 0.0  & 0.0  & 0.0  & 0.0  & 0.000  & 0.4671  & 1544  \\ \hline
        52 & RIT:eBBH:1560 & Nonspinning & 24.59  & 7  & 0.0  & 0.0  & 0.0  & 0.0  & 0.0  & 0.0  & 0.000  & 0.1900  & 22937  \\ \hline
        53 & RIT:eBBH:1561 & Nonspinning & 24.59  & 7  & 0.0  & 0.0  & 0.0  & 0.0  & 0.0  & 0.0  & 0.000  & 0.3600  & 6491  \\ \hline
        54 & RIT:eBBH:1562 & Nonspinning & 24.59  & 7  & 0.0  & 0.0  & 0.0  & 0.0  & 0.0  & 0.0  & 0.000  & 0.4375  & 2680  \\ \hline
        55 & RIT:eBBH:1563 & Nonspinning & 24.59  & 7  & 0.0  & 0.0  & 0.0  & 0.0  & 0.0  & 0.0  & 0.000  & 0.4671  & 1733 \\ \hline
    \end{tabular}
\end{table*}

\begin{table*}[htbp]
\caption{\label{tab:II}Eccentric and circular orbit waveforms from the SXS catalog, which contains simulation number, type, mass ratio $q\in[1,7]$, initial separation $\operatorname{sep}_{ini}$, dimensionless spin component, effective spin, initial eccentricity $e_0$ after removing junk radiation, merge time $t_{merger}$ and   orbit number $N_{orbits}$.}
    \centering
    \begin{tabular}{|r|r|r|r|r|r|r|r|r|r|r|r|r|r|r|}
    \hline
        case & simulations & type & q & $sep_{ini}$ & $\chi_{1x}$ & $\chi_{1y}$ & $\chi_{1z}$ & $\chi_{2x}$ & $\chi_{2y}$ & $\chi_{2z}$ & $\chi_{eff}$ & $e_0$ & $t_{merge}/M$ & $N_{orbits}$\\ \hline
        2 & SXS:BBH:1355 & Nonspinning & 1  & 12.98  & 0.0  & 0.0  & 0.0  & 0.0  & 0.0  & 0.0  & 0.0  & 0.0613  & 2552  & 13.9 \\ \hline
        3 & SXS:BBH:1356 & Nonspinning & 1  & 18.87  & 0.0  & 0.0  & 0.0  & 0.0  & 0.0  & 0.0  & 0.0  & 0.1010  & 6001  & 22.34 \\ \hline
        4 & SXS:BBH:1357 & Nonspinning & 1  & 16.24  & 0.0  & 0.0  & 0.0  & 0.0  & 0.0  & 0.0  & 0.0  & 0.1123  & 2889  & 14.76 \\ \hline
        5 & SXS:BBH:1358 & Nonspinning & 1  & 15.90  & 0.0  & 0.0  & 0.0  & 0.0  & 0.0  & 0.0  & 0.0  & 0.1097  & 2656  & 14.08 \\ \hline
        6 & SXS:BBH:1359 & Nonspinning & 1  & 15.74  & 0.0  & 0.0  & 0.0  & 0.0  & 0.0  & 0.0  & 0.0  & 0.1103  & 2530  & 13.75 \\ \hline
        7 & SXS:BBH:1360 & Nonspinning & 1  & 16.70  & 0.0  & 0.0  & 0.0  & 0.0  & 0.0  & 0.0  & 0.0  & 0.1643  & 2373  & 13.14 \\ \hline
        8 & SXS:BBH:1361 & Nonspinning & 1  & 16.70  & 0.0  & 0.0  & 0.0  & 0.0  & 0.0  & 0.0  & 0.0  & 0.1691  & 2325  & 12.98 \\ \hline
        9 & SXS:BBH:1362 & Nonspinning & 1  & 17.70  & 0.0  & 0.0  & 0.0  & 0.0  & 0.0  & 0.0  & 0.0  & 0.2195  & 2147  & 12.28 \\ \hline
        10 & SXS:BBH:1363 & Nonspinning & 1  & 17.70  & 0.0  & 0.0  & 0.0  & 0.0  & 0.0  & 0.0  & 0.0  & 0.2208  & 2109  & 12.18 \\ \hline
        11 & SXS:BBH:0154 & Spin-aligned & 1  & 15.37  & 0.0  & 0.0  & -0.8  & 0.0  & 0.0  & -0.8  & -0.8  & 0.0000  & 3805  & 13.24 \\ \hline
        12 & SXS:BBH:0325 & Spin-aligned & 1  & 15.87  & 0.0  & 0.0  & 0.0  & 0.0  & 0.0  & -0.8  & -0.4  & 0.0000  & 5104  & 17.34 \\ \hline
        13 & SXS:BBH:1165 & Nonspinning & 2  & 19.92  & 0.0  & 0.0  & 0.0  & 0.0  & 0.0  & 0.0  & 0.0  & 0.0000  & 15792  & 40.34 \\ \hline
        14 & SXS:BBH:1364 & Nonspinning & 2  & 14.54  & 0.0  & 0.0  & 0.0  & 0.0  & 0.0  & 0.0  & 0.0  & 0.0522  & 3200  & 16.14 \\ \hline
        15 & SXS:BBH:1365 & Nonspinning & 2  & 15.03  & 0.0  & 0.0  & 0.0  & 0.0  & 0.0  & 0.0  & 0.0  & 0.0677  & 3181  & 16.06 \\ \hline
        16 & SXS:BBH:1366 & Nonspinning & 2  & 16.03  & 0.0  & 0.0  & 0.0  & 0.0  & 0.0  & 0.0  & 0.0  & 0.1110  & 3073  & 15.62 \\ \hline
        17 & SXS:BBH:1367 & Nonspinning & 2  & 15.88  & 0.0  & 0.0  & 0.0  & 0.0  & 0.0  & 0.0  & 0.0  & 0.1091  & 2955  & 15.31 \\ \hline
        18 & SXS:BBH:1368 & Nonspinning & 2  & 15.73  & 0.0  & 0.0  & 0.0  & 0.0  & 0.0  & 0.0  & 0.0  & 0.1066  & 2850  & 15.03 \\ \hline
        19 & SXS:BBH:1369 & Nonspinning & 2  & 18.18  & 0.0  & 0.0  & 0.0  & 0.0  & 0.0  & 0.0  & 0.0  & 0.2220  & 2616  & 13.92 \\ \hline
        20 & SXS:BBH:1370 & Nonspinning & 2  & 17.70  & 0.0  & 0.0  & 0.0  & 0.0  & 0.0  & 0.0  & 0.0  & 0.2167  & 2377  & 13.23 \\ \hline
        21 & SXS:BBH:2265 & Nonspinning & 3  & 22.56  & 0.0  & 0.0  & 0.0  & 0.0  & 0.0  & 0.0  & 0.0  & 0.0000  & 30923  & 65.56 \\ \hline
        22 & SXS:BBH:1371 & Nonspinning & 3  & 14.97  & 0.0  & 0.0  & 0.0  & 0.0  & 0.0  & 0.0  & 0.0  & 0.0627  & 3707  & 18.17 \\ \hline
        23 & SXS:BBH:1372 & Nonspinning & 3  & 15.97  & 0.0  & 0.0  & 0.0  & 0.0  & 0.0  & 0.0  & 0.0  & 0.1064  & 3565  & 17.66 \\ \hline
        24 & SXS:BBH:1373 & Nonspinning & 3  & 15.84  & 0.0  & 0.0  & 0.0  & 0.0  & 0.0  & 0.0  & 0.0  & 0.1058  & 3451  & 17.31 \\ \hline
        25 & SXS:BBH:1374 & Nonspinning & 3  & 18.10  & 0.0  & 0.0  & 0.0  & 0.0  & 0.0  & 0.0  & 0.0  & 0.2264  & 3015  & 15.59 \\ \hline
        26 & SXS:BBH:1220 & Nonspinning & 4  & 15.20  & 0.0  & 0.0  & 0.0  & 0.0  & 0.0  & 0.0  & 0.0  & 0.0000  & 7301  & 26.26 \\ \hline
        27 & SXS:BBH:0056 & Nonspinning & 5  & 15.00  & 0.0  & 0.0  & 0.0  & 0.0  & 0.0  & 0.0  & 0.0  & 0.0000  & 7980  & 28.81 \\ \hline
        28 & SXS:BBH:0181 & Nonspinning & 6  & 14.00  & 0.0  & 0.0  & 0.0  & 0.0  & 0.0  & 0.0  & 0.0  & 0.0000  & 6695  & 26.47 \\ \hline
        29 & SXS:BBH:0298 & Nonspinning & 7  & 12.20  & 0.0  & 0.0  & 0.0  & 0.0  & 0.0  & 0.0  & 0.0  & 0.0000  & 4251  & 19.68 \\ \hline
    \end{tabular}
\end{table*}

\end{document}